\documentclass{article}

\usepackage{amsmath}
\usepackage{amsfonts}
\usepackage{amssymb}
\usepackage{latexsym}
\usepackage{stmaryrd}
\usepackage{array}
\usepackage{exscale}
\newcommand{\nc}{\newcommand}
\newcommand{\ol}{\overline}

\newcommand{\es}{\emptyset}
\newcommand{\sm}{\setminus}
\newcommand{\ve}{\varepsilon}
\newcommand{\vp}{\varphi}

\newcommand{\bc}{\bigcup}
\newcommand{\bca}{\bigcap}
\newcommand{\Lra}{\Leftrightarrow}

\newcommand{\Ra}{\Rightarrow}

\newcommand{\ra}{\rightarrow}

\newcommand{\sse}{\subseteq}

\newcommand{\fa}{\forall}
\newcommand{\ex}{\exists}
\newcommand{\mr}{\mathrm}
\newcommand{\mc}{\mathcal}

\newcommand{\DMO}{\DeclareMathOperator}
\newcommand{\DST}{\displaystyle}

\newcommand{\NN}{\mathbb{N}}
\newcommand{\NNZ}{\NN_0}

\newcommand{\und}{{\:\wedge\:}} 
\newcommand{\mb}{{\:|\:}} 
\newcommand{\set}[1]{\{ #1 \}}
\newcommand{\setb}[1]{\big \{ \, #1 \, \big \}}
\DeclareMathOperator{\dom}{dom}

\nc{\simlvi}[1]{\!\sim_{#1}}
\newcommand{\rbca}[1]{\bca_{#1}\nolimits} 

\DeclareMathOperator{\addcup}{{\stackrel{\text{\raisebox{-2.2ex}[-0ex][-0ex]{\large$\cdot$}}}{\cup}}} 
\nc{\apprel}[3]{{#1}(#2)_{(#3)}} 
\newcommand{\nni}{\NNZ \cup \{+\infty\}} 
\providecommand{\abs}[1]{\lvert #1 \rvert} 
\newcommand{\Va}{\mc{V\hspace{-0.1em}A}}

\newcommand{\Lit}{\mc{LIT}}
\newcommand{\Cl}{\mc{CL}}
\newcommand{\Cls}{\mc{CLS}}

\newcommand{\Pcls}[1]{#1\mbox{--}\Cls}

\newcommand{\Pass}{\mc{P\hspace{-0.32em}ASS}}
\newcommand{\epa}{\pab{}} 
\newcommand{\Tass}{\mc{T\hspace{-0.35em}ASS}}
\newcommand{\Sat}{\mc{SAT}}

\newcommand{\Usat}{\mc{USAT}}

\nc{\Clsoo}{\Cls^{1,1}} 
\DeclareMathOperator{\lit}{lit}
\DeclareMathOperator{\var}{var}

\newcommand{\Ho}{\mc{HO}} 
\newcommand{\Rho}{\mc{R}\Ho} 
\newcommand{\Qho}{\mc{Q}\Ho}
\newcommand{\Dnf}{\mathrm{DNF}}
\newcommand{\Cnf}{\mathrm{CNF}}
\DeclareMathOperator{\res}{\diamond} 

\DeclareMathOperator{\comp}{Comp} 
\DeclareMathOperator{\comptr}{\comp_{R}^*} 

\DeclareMathOperator{\hardness}{hd}
\DMO{\phardness}{phd} 
\DMO{\hts}{hs} 
\newcommand{\php}{\mathrm{PHP}}

\newcommand{\pab}[1]{\langle #1 \rangle}
\newcommand{\pao}[2]{\langle #1 \ra #2 \rangle}

\DMO{\rsub}{r_S} 
\DMO{\rk}{r} 
\DMO{\rki}{r_{\infty}} 
\nc{\rslur}{\xrightarrow{\text{SLUR}}} 
\nc{\rslurs}{\rslur_{\!*}} 
\DMO{\slur}{slur} 
\nc{\Slur}{\mc{SLUR}} 
\nc{\rkslur}[1]{\xrightarrow{\text{SLUR}_{#1}}} 
\nc{\rkslurs}[1]{\rkslur{#1}_{\!*}} 
\nc{\Urefc}{\mc{UC}}
\nc{\Propc}{\mc{PC}}
\DeclareMathOperator{\primec}{prc} 
\DMO{\frl}{fl} 
\usepackage{theorem} 
\usepackage[driverfallback=hypertex]{hyperref}
\newtheorem{defi}{Definition}[section]
\newtheorem{lem}[defi]{Lemma}
\newtheorem{thm}[defi]{Theorem}
\newtheorem{corol}[defi]{Corollary}

\newtheorem{conj}[defi]{Conjecture}

\newtheorem{examp}[defi]{Example}

\theorembodyfont{\rmfamily}

\theorembodyfont{}
\newenvironment{prf}{\noindent\textbf{Proof:}\;}{\par\noindent\ignorespacesafterend}

\newcommand{\Qed}{\hfill $\square$}
\nc{\bm}{\boldmath}
\nc{\bmm}[1]{\mbox{\bm$\DST #1$}}
\nc{\mi}[1]{\bmm{\mathrm{(#1):}} \quad}

\usepackage[all]{xy}
\usepackage{enumerate}
\usepackage{xr}
\usepackage[longnamesfirst,numbers]{natbib}
\usepackage{array}

\renewcommand{\cite}{\citet}

\nc{\implk}[1]{\models_{#1}} 
\DMO{\whardness}{whd} 

\nc{\Altsluri}[1]{\Slur(#1)}
\nc{\Altslurstar}{\Slur\text{\textasteriskcentered}}
\nc{\Altslurstari}[1]{\Altslurstar(#1)}
\nc{\Canon}{\mr{CANON}}
\nc{\Canoni}[1]{\Canon(#1)}
\nc{\rkslurstar}[1]{\xrightarrow{\text{SLUR\textasteriskcentered}#1}} 
\nc{\rkslursstar}[1]{\rkslurstar{#1}_{\!*}} 
\DMO{\slurstar}{\slur\!\text{\textasteriskcentered}}
\DMO{\slurstarrelop}{R}
\nc{\slurstarrel}[1]{\mathrel{\slurstarrelop_{#1}}}

\begin{document}

\title{Generalising unit-refutation completeness and SLUR via nested input resolution}

\author{
  Matthew Gwynne\\
  Computer Science Department\\
  College of Science, Swansea University\\
  Swansea, SA2 8PP, UK\\
  email: csmg@swansea.ac.uk\\
  {\small \url{http://cs.swan.ac.uk/~csmg/}}
  \and Oliver Kullmann\\
  Computer Science Department\\
  College of Science, Swansea University\\
  Swansea, SA2 8PP, UK\\
  email: O.Kullmann@Swansea.ac.uk\\
  {\small \url{http://cs.swan.ac.uk/~csoliver}}
 }

\maketitle

\begin{abstract}
  The class $\Slur$ (\emph{Single Lookahead Unit Resolution}) was introduced in \cite{SAFS95} as an umbrella class for efficient SAT solving, with in fact linear time SAT decision (while the recognition problem was not considered). \cite{CepekKuceraVlcek2012SLUR,BalyoGurskyKuceraVlcek2012SLURHier} extended this class in various ways to hierarchies covering all of CNF (all clause-sets). We introduce a hierarchy $\Slur_k$ which we argue is the natural ``limit'' of such approaches.

  The second source for our investigations is the class $\Urefc$ of \emph{unit-re\-fu\-ta\-tion complete clause-sets} introduced in \cite{Val1994UnitResolutionComplete}. Via the theory of (tree-resolution based) ``hardness'' of clause-sets as developed in \cite{Ku99b,Ku00g,AnsoteguiBonetLevyManya2008Hardness} we obtain a natural generalisation $\Urefc_k$, containing those clause-sets which are ``unit-refutation complete of level $k$'', which is the same as having hardness at most $k$. Utilising the strong connections to (tree-)resolution complexity and (nested) input resolution, we develop fundamental methods for the determination of hardness (the level $k$ in $\Urefc_k$).

  A fundamental insight now is that $\Slur_k = \Urefc_k$ holds for all $k$. We can thus exploit both streams of intuitions and methods for the investigations of these hierarchies. As an application we can easily show that the hierarchies from \citep{CepekKuceraVlcek2012SLUR,BalyoGurskyKuceraVlcek2012SLURHier} are strongly subsumed by $\Slur_k$.

 Finally we consider the problem of ``irredundant'' clause-sets in $\Urefc_k$. For 2-CNF we show that strong minimisations are possible in polynomial time, while already for (very special) Horn clause-sets minimisation is NP-complete. We conclude with an extensive discussion of open problems and future directions.
\end{abstract}

\tableofcontents

\section{Introduction}
\label{sec:intro}

The boolean satisfiability problem, SAT for short, in its core version is the problem of deciding satisfiability of a conjunctive normal form (clause-set) $F$; see the handbook \cite{2008HandbuchSAT} for further information. An important theme is the the search for relevant classes $\mc{C}$ of clause-sets $F$ for which one can (at least) decide satisfiability in \emph{polynomial time} (that is, deciding whether $F$ logically implies the empty clause); see Section 1.19 in \cite{FM09HBSAT} for some basic information. For the task of \emph{knowledge compilation} one wants more from the target-class $\mc{C}$, namely that the clausal entailment problem (deciding whether $F$ logically implies some given clause) can be decided in polynomial time; see \cite{DarwicheMarquis2002KCmap} for an overview. In this report now we bring together two previously unconnected streams of research from these two areas:
\begin{description}
\item[SLUR] The SLUR algorithm is an incomplete linear-time SAT-decision algorithm, based on look-ahead via unit-clause propagation.
\item[UC] The class UC of unit-refutation complete clause-sets enables clausal-en\-tail\-ment decision in linear time via unit-clause propagation.
\end{description}
In Subsections \ref{sec:introslurh}, \ref{sec:introhd} we will discuss these two streams in turn, while their unification is outlined in Subsection \ref{sec:introtogether}, and applications to ``SAT knowledge compilation'' are discussed in Subsection \ref{sec:introoutlook}. This is the underlying report of the conference-version \cite{GwynneKullmann2012SlurSOFSEM}, while the journal-version is \cite{GwynneKullmann2012SlurJ}.

\subsection{The quest for SLUR hierarchies}
\label{sec:introslurh}

In the year 1995 in \cite{SAFS95} the SLUR algorithm was introduced, a simple incomplete non-deterministic SAT-decision algorithm, which always succeeded on various classes with polynomial-time SAT decision where previously only rather complicated algorithms were known. The computation is divided into two phases for input-clause-set $F$: First we check via unit-clause propagation (UCP) for unsatisfiability. If this check fails, then we assume $F$ is satisfiable, and guess a satisfying assignment, using UCP-look-ahead for the guessed assignments to avoid obviously false assignments. The class $\Slur$ contains those $F$ where this algorithm always succeeds (i.e., always finds a satisfying assignment in the second phase).

So recognition of $\Slur$ seems a non-trivial problem, while SAT decision for $F \in \Slur$ can be done in linear time. The natural question arises, whether $\Slur$ can be turned into a hierarchy, covering in the limit all clause-sets. A generalisation of SLUR has been considered in \cite{FrancoSchlipf1997Report} under the name ``ISLUR'' (improved SLUR), allowing a polynomial number $p(\ell(F))$ of backtracks (for a fixed polynomial $p$, in the input-size $\ell(F)$), in the unsatisfiability as well as in the satisfiability phase of the SLUR algorithm, before giving up. It is mentioned that ISLUR gives up on every large enough ``sparse'' clause-set (which are ``typical'' as random k-CNF clause-sets), when no variable occurs ``too often''. This was considered to be ``disappointing'' --- but from our point of view the value of the class $\Slur$ lies not in being a ``big'' class of clause-sets with polynomial-time SAT solving, but in establishing a basic target class for representations of boolean functions with very strong properties via clause-sets; see Subsection \ref{sec:introoutlook} for further discussions. For all fixed $k$ there exists a polynomial $p$ such the $k$-th level of our hierarchy, $\Slur_k$, is contained in the class ISLUR (those clause-sets where the ISLUR algorithm never gives up). So all levels are negligible when considering the above sparse clause-sets, but as we will argue in Subsection \ref{sec:introoutlook}, nevertheless this hierarchy is proper regarding good representations of boolean functions, and the parameter $k$ is meaningful and robust (not just a numerical parameter like the polynomial $p$).

In \cite{CepekKuceraVlcek2012SLUR,BalyoGurskyKuceraVlcek2012SLURHier} the authors finally proved that membership decision of $\Slur$ is coNP-complete, and presented three hierarchies, $\Altsluri{k}, \Altslurstari{k}$ and $\Canoni{k}$. It still seemed that none of these hierarchies is the final answer, though they all introduce a certain natural intuition. We now present what seems the natural ``limit hierarchy'', which we call $\Slur_k$, and which unifies the two basic intuitions embodied in $\Altsluri{k}, \Altslurstari{k}$ on the one hand and $\Canoni{k}$ on the other hand.

In order to do so we need a precise analysis of the $\Slur$-class. We introduce the SLUR transition relation $F \rslur F'$ between clause-sets $F, F'$, which makes precise one non-deterministic step of the SLUR-algorithm. This transition from $F$ to $F'$ happens when assigning a (single) literal in such a way that UCP does not create the empty clause. The core of the classes $\Altsluri{k}$ and $\Altslurstari{k}$ is to strengthen the transition relation by requesting that not just one literal is choosable, but actually $k$ literals can be chosen, while the difference between them is that $\Altslurstari{k}$ performs UCP inbetween the choices, while the weaker class $\Altsluri{k}$ does not.

Before we can describe our solution, the $\Slur_k$-hierarchy, we need to discuss the second source of our approach, the class $\Urefc$ of ``unit-refutation complete clause-sets'', which is related to the stream embodied by $\Canoni{k}$.

\subsection{Unit-refutation completeness and ``hardness''}
\label{sec:introhd}

In the year 1994 in \cite{Val1994UnitResolutionComplete} the class $\Urefc$ was introduced, containing clause-sets $F$ such that clausal entailment, that is, whether $F \models C$ holds (clause $C$ follows logically from $F$, i.e., $C$ is an implicate of $F$), can be decided by unit-clause propagation. The motivation was knowledge compilation, that is, to have a more succinct alternative to the use of the set of all prime implicates of a given clause-set $F_0$ (clausal database), for which one seeks an equivalent $F$ such that clausal entailment can be decided quickly.

A second development is important here, namely the development of the notion of ``hardness'' in \cite{Ku99b,Ku00g,AnsoteguiBonetLevyManya2008Hardness}. The first source \citep{Ku99b} from 1999 introduced the notion of hardness as a measure $\hardness_0: \Cls \ra \NNZ$, assigning natural numbers to clause-sets in the following way (using $\Sat \subset \Cls$ for the satisfiable clause-sets, and $\Usat := \Cls \sm \Sat$):
\begin{itemize}
\item $\hardness_0(F) := 0$ for the simplest clause-sets $F \in \Cls$ regarding SAT decision, containing the empty clause (i.e., $\bot \in F$) or being empty (i.e., $F = \top$).\footnote{Actually a two-dimensional family $\hardness_{\mc{U},\mc{S}}$ of such measures was introduced, based on oracles $\mc{U} \sse \Usat$, $\mc{S} \sse \Sat$ for deciding unsatisfiability resp.\ satisfiability, and setting $\hardness_{\mc{U},\mc{S}}(F) := 0$ for $F \in \mc{U} \cup \mc{S}$. In this report we consider only the simplest base case $\hardness_0 = \hardness_{\mc{U}_0,\mc{S}_0}$, where $\mc{U}_0 := \set{F \in \Cls : \bot \in F}$ and $\mc{S} := \set{\top}$. Oracle $\mc{S}$ does not play a role in the setting of this report, which is fully unsatisfiability-based. See Subsection \ref{sec:Alternativehierarchies} for more information on these hierarchies, and see Subsection \ref{sec:outlinerelhd} for an outlook on relativised hardness.}
\item $\hardness_0(F) = k \ge 1$ iff there is a literal $x$ such that for $F' := \pao x0 * F$ (setting $x$ to $0$) we have $\hardness_0(F') \le k-1$ and either $F' \in \Usat$ and $\hardness_0(\pao x1 * F) \le k$, or $F' \in \Sat$.
\end{itemize}
The second source \citep{Ku00g} from 2004 generalised this approach to constraint satisfaction problems (and beyond). The third source \citep{AnsoteguiBonetLevyManya2008Hardness} from 2008 considered $\hardness_0(F)$ on unsatisfiable clause-sets $F \in \Usat$, relating it to backdoors, cycle-cutsets and treewidth, and performing an experimental study on random instances. Also in \citep{AnsoteguiBonetLevyManya2008Hardness} we find a different extension of $\hardness_0: \Usat \ra \NNZ$ to a measure $\hardness: \Cls \ra \NNZ$, using for satisfiable instances $F \in \Sat$ the maximisation over all unsatisfiable sub-instances obtained by applying partial assignments. This hardness notion is harder to measure: as we show in this report, determining whether $\hardness(F) \le k$ holds for a fixed $k \ge 1$ is coNP-complete, while $\hardness_0(F) \le k$ can be decided in polynomial time (for fixed $k$). Nevertheless it is the central measure for this report, and we consider it as measuring ``representation hardness'', while $\hardness_0$ measures ``solver hardness''.\footnote{$\hardness(F)$ actually captures tree-like resolution (in a sense). In Subsection \ref{sec:outlinewhd} we discuss a width-based measure of hardness, which captures dag-like resolution. We consider the tree-hardness as the natural starting point.}

As we show in Theorem \ref{thm:characuckir}, $\hardness(F) \le k$ is equivalent to the property of $F$, that all implicates of $F$ (i.e., all clauses $C$ with $F \models C$) can be derived by $k$-times nested input resolution from $F$, a generalisation of input resolution as introduced and studied in \citep{Ku99b,Ku00g}.\footnote{Equivalently, as shown in \citep{Ku99b,Ku00g}, one can say that all implicates $C$ have a tree-resolution proof using space at most $k+1$.} So we obtain that $\Urefc$ is precisely the class of clause-sets $F$ with $\hardness(F) \le 1$ ! It is then natural to define the hierarchy $\Urefc_k$ via the property $\hardness(F) \le k$. The hierarchy $\Canoni{k}$ is based on resolution trees of height at most $k$, which is a special case of $k$-times nested input resolution, and so we have $\Canoni{k} \subset \Urefc_k$.

\subsection{Bringing SLUR and UC together}
\label{sec:introtogether}

In order to get back to SLUR, we need to emphasise the two-sided nature of the hardness measure, as developed in \citep{Ku99b,Ku00g}. In Subsection \ref{sec:introhd} we discussed the \emph{proof-theoretic side} of it. The \emph{algorithmic side} is given by the reductions $\rk_k: \Cls \ra \Cls$ (introduced in \citep{Ku99b}), which perform certain forced assignments:
\begin{enumerate}
\item $\rk_1$ is UCP, assigning $x \ra 1$ for unit-clauses $\set{x}$ until all are eliminated.
\item $\rk_2$ is (complete) failed-literal elimination, assigning, while possible, $x \ra 1$ for literals $x$ such that the assignment $x \ra 0$ yields a contradiction via $\rk_1$; see Section 5.2.1 in \cite{HvM09HBSAT} for the usage of failed literals in SAT solvers (so-called ``look-ahead solvers''), and see Section 7.2.2 in \cite{Kullmann2007HandbuchTau} for the general explanation of $\rk_2$ being the ``look-ahead version'' of $\rk_1$.
\item In general $\rk_{k+1}$ is the ``look-ahead version'' of $\rk_k$, assigning, while possible, $x \ra 1$ for literals $x$ such that the assignment $x \ra 0$ yields a contradiction via $\rk_k$.
\end{enumerate}
For unsatisfiable $F$ the hardness $\hardness(F)$ is equal to the minimal $k$ such that $\rk_k(F)$ detects unsatisfiability of $F$, i.e., $\rk_k(F) = \set{\bot}$. This yields the basic observation $\Urefc \sse \Slur$ --- and actually we have $\Urefc = \Slur$ !

So by replacing the use of $\rk_1$ in the SLUR algorithm by $\rk_k$ (using our analysis via the transition relation) we obtain a natural hierarchy $\Slur_k$, which includes the previous SLUR-hierarchies $\Altsluri{k}$ and $\Altslurstari{k}$, and where we have $\Slur_k = \Urefc_k$. This equality of these two hierarchies is our argument that we have found the ``limit hierarchy'' for SLUR.

\subsection{Outlook on good representations of boolean functions}
\label{sec:introoutlook}

The ideas presented in Subsections \ref{sec:introslurh} to Subsection \ref{sec:introtogether} are the main thrust for the results of this paper (Sections \ref{sec:SLURh} to \ref{sec:slurhier}), while in the final Section \ref{sec:goodrpw} (and also in the outlook in Section \ref{sec:conclusion}) we touch upon what we consider as the main application area and the main area for future developments of the theory, namely a theory of good representations of boolean functions. More precisely, in Section \ref{sec:goodrpw} we consider the complexity of finding short equivalent clause-sets of bounded hardness for the most basic CNF classes, 2-CNF and Horn clause-sets, and we show feasibility for the former, NP-completeness for the latter. We roughly outline now the basic ideas on ``good representations'' in general, while in Section \ref{sec:conclusion} some more details are presented.

SAT algorithms have seen an astounding development in the last two decades. Especially efficient algorithms, data structures and heuristics have been developed. The main bottleneck currently is that the underlying constraint problem needs to be represented via boolean CNF, and it is not clear at all how to do this so that SAT solving becomes as easy as possible. ``SAT modulo Theories'' (SMT; see \cite{BSST09HBSAT}) boosts the representation by extending the general method, however it does not yield insights into how to construct the basic representations by CNFs. What is needed is a systematic investigation into ``good representations'' of boolean functions $f$ by clause-sets $F$, with the aim of ``intelligent'' SAT translations.

As a first answer, we consider the classes $\Urefc_k$ as the most basic target classes, that is, $F \in \Urefc_k$ for $k$ ``as small as possible'' is the (basic) fundamental guideline. The motivation for $\Urefc$ was that of a ``good representation'', while the motivation for $\Slur$ was ``good SAT solving'' --- the hierarchies $\Urefc_k = \Slur_k$ bring these two aspects together, and this in a parameterised way, so that $k$ can be traded against the size of $F$. So the theory of good representations $F$ of boolean functions $f$ can be considered as ``SAT knowledge representation'', where the ``knowledge'', the boolean function $f$, must be represented by a clause-set $F$ such that all ``aspects'' of $f$ (most fundamental the prime implicates) are represented in such a way that a SAT solver can ``understand'' this representation.

What is now the precise relation between the boolean function $f$ to be represented, and the representation $F$, a clause-set? The most basic idea is to consider that $F$ as a CNF is equivalent to $f$, which we write as $F \cong f$ (more precisely, $\Cnf(F) \cong f$). Good representations in this (restricted) setting then amount to consider subsets $F \sse \primec_0(f)$ of the set of prime implicates of $f$, such that $F \cong f$ and such that $\hardness(F)$ and $\ell(F)$ (the size of $F$) are in a ``reasonable'' relationship (the lower $\hardness(F)$ the higher $\ell(F)$, and so a balance is to be sought). The basic conjecture then states that allowing larger hardness yields more possibilities for short representations:
\begin{conj}\label{con:hierarchygoodrepw}
  For every $k \in \NNZ$ there exists a sequence $(f_n)_{n \in \NN}$ of boolean functions, such that no polysize-sequence $(F_n)_{n \in \NN}$ (i.e., where $(\ell(F_n))_{n \in \NN}$ is polynomially bounded in $n$) exists with
  \begin{itemize}
  \item $F_n \cong f_n$
  \item $\hardness(F_n) \le k$
  \end{itemize}
  for all $n$, but where such a sequence $(F_n)_{n \in \NN}$ exists when allowing $\hardness(F_n) \le k+1$.
\end{conj}
Conjecture \ref{con:nocollapseabshd} extends this conjecture to include the use of new variables, and also refines it by introducing intermediate levels between the hardness-levels.\footnote{In \cite{GwynneKullmann2013GoodRepresentations} we have meanwhile established that Conjecture \ref{con:hierarchygoodrepw} is true.}

The algorithmic approach for such representations (not using new variables) is to systematically search for small $F$ with a given hardness upper-bound. In Section \ref{sec:goodrpw} one finds the most basic considerations. In \cite{GwynneKullmann2011TranslationsPrelim} we presented some initial experimental results on using this approach for the (small) building-blocks like the S-boxes in block ciphers like AES and DES, for their SAT-based cryptanalysis (see Subsection \ref{sec:outlinecrypt} for more information).

\subsection{The Schaefer classes}
\label{sec:Schaeffer}

We conclude by some remarks on the four main classes from Schaefer's dichotomy result (see Section 12.2 in \cite{DH09HBSAT} for an introduction, and see \cite{CreignouKolaitisVollmer2008ComplexityConstraints} for an in-depth overview on recent developments). Our point of view here is that we consider a boolean function $f$ which is either Horn, dual Horn, bijunctive or affine, and we ask for a good representation $F \in \Cls$ of $f$:
\begin{itemize}
\item If $f$ is Horn or dual Horn, then there is a (dual) Horn clause-set $F$ equivalent to $f$, and by Part \ref{lem:hdupperb4} of Lemma \ref{lem:hdupperb} we have $\hardness(F) \le 1$. So obtaining a representation $F \in \Urefc$ is trivial; however optimising the size of $F$ is NP-complete (see Theorem \ref{thm:NPcomp1baseHO}).
\item If $f$ is bijunctive, then there is a 2-CNF $F$ equivalent to $f$, and by Part \ref{lem:hdupperb3} of Lemma \ref{lem:hdupperb} we have $\hardness(F) \le 2$. Moreover, by Theorem \ref{thm:opt2cnfw} we can reduce the hardness to $0$ or $1$ (as we wish) in polynomial time, and that by optimal (shortest) such $F$.
\item If $f$ is affine, that is, $f$ is the conjunction of $m$ linear equations $x_1 \oplus \dots \oplus x_p = 0$ over $\set{0,1}$ viewed as a 2-element field, with addition $\oplus$ as exclusive-or, then the situation regarding the existence of a representation of bounded hardness is not fully understood yet:
  \begin{enumerate}
  \item If $m=1$, then there is precisely one CNF-representation of $f$ without new variables, containing $2^{p-1}$ clauses and being (trivially) of hardness $0$. So without new variables we have a polysize representation of bounded hardness iff $p$ is bounded.
  \item While when allowing new variables, then for $m=1$ there is a representation $F \in \Urefc$, as will be shown in \cite{GwynneKullmann2013GoodRepresentations}.
  \item For arbitrary $m$ there is definitely no small representation without new variables when the clause-length $p$ is unbounded. When bounding $p$, or when allowing new variables, then the existence of a polysize $F \in \Urefc_k$ for some fixed $k$ seems to be an interesting open problem; for some partial results see \cite{LaitinenJunttilaNiemelae2012Parity}. Perhaps no polysize representations $F \in \Urefc$ exist, even for the ``relative condition'', where propagation-conditions are posed only for the variables in the XOR-clauses; see \cite{BKNW2009CircuitComplexity} for general tools for such lower bounds, and see Subsections \ref{sec:outlinegoodrep}, \ref{sec:outlinerelhd} for more discussions.
  \end{enumerate}
\end{itemize}

\subsection{Overview}
\label{sec:overview}

After discussing basic terminology in \textbf{Section \ref{sec:prelim}}, in \textbf{Section \ref{sec:SLURh}} we discuss SLUR and existing extensions. We give a precise (mathematical) definition of the class $\Slur$, achieving a conceptually clear understanding, and based on these concepts we give precise (mathematical) definitions of the various SLUR hierarchies from the literature. In \textbf{Section \ref{sec:genucp}} we provide the background about generalised unit-clause propagation, that is, the reductions $\rk_k: \Cls \ra \Cls$, where $\Cls$ is the set of all clause-sets and $\rk_1$ is unit-clause propagation. \textbf{Section \ref{sec:Hardness}} then introduces the hardness $\hardness: \Cls \ra \NNZ$ and defines the classes $\Urefc_k \subset \Cls$ of ``unit-refutation complete clause-sets of level $k$'' as those $F$ with $\hardness(F) \le k$. The first main result is \textbf{Theorem \ref{thm:characuckir}}, which states that the elements of $\Urefc_k$ are precisely the clause-sets $F$ where every prime implicate of $F$ can be derived by $k$-times nested input resolution from $F$. In Section \ref{sec:fundprop} we develop various tools to determine hardness. First we consider various constructions in Subsection \ref{sec:basichd}. Then in Subsection \ref{sec:uckcontain} we provide tools to show that classes of clause-sets have bounded hardness, with applications to common classes and to stability properties of the classes $\Urefc_k$. Alternative and generalised hardness-notions are considered in Subsection \ref{sec:Alternativehierarchies}. We conclude by considering algorithmic ways to determine the hardness-measure in Subsection \ref{sec:Dethd}. Section \ref{sec:slurhier} introduces the $\Slur_k$ hierarchy. Our second major result is \textbf{Theorem \ref{thm:slurhdk}}, showing that $\Urefc_k = \Slur_k$ holds. From this characterisation we derive in \textbf{Theorem \ref{thm:dethdconpcp}} the coNP-completeness of membership decision for $\Urefc_k$ when $k \ge 1$. And in Theorems \ref{thm:altcanonweak}, \ref{thm:altslurweak} we show that the previous hierarchies are (strictly) included in the $\Slur_k$ hierarchy, which we consider as a kind of ``completion'', where both approaches, based on SLUR and UC, meet. In \textbf{Section \ref{sec:goodrpw}} we turn towards the problem of finding short equivalent clause-sets of low hardness for a given clause-set $F$. In \textbf{Theorem \ref{thm:opt2cnfw}} we show that for $F$ in 2-CNF we can compute optimal equivalent clause-sets (of low hardness) in polynomial time. While in \textbf{Theorem \ref{thm:NPcomp1baseHO}} we show that already for Horn clause-sets $F$, even when all prime implicates are given as part of the input, the decision whether there is an equivalent clause-set (of low hardness) using at most a given number of clauses is NP-complete. We conclude in \textbf{Section \ref{sec:conclusion}} with the summary and an extensive discussion of future directions.

\section{Preliminaries}
\label{sec:prelim}

We follow the general notions and notations as outlined in \cite{Kullmann2007HandbuchMU}. We use $\NN = \set{1,\dots}$ and $\NNZ = \NN \cup \set{0}$. Based on an infinite set $\Va$ of variables, we form the set $\Lit := \Va \addcup \ol{\Va}$ of positive and negative literals, using complementation. A clause $C \subset \Lit$ is a finite set of literals without clashes, i.e., $C \cap \ol{C} = \es$, where for $L \sse \Lit$ we set $\ol{L} := \set{\ol{x} : x \in L}$. The set of all clauses is denoted by $\Cl$. A clause-set $F \subset \Cl$ is a finite set of clauses, and the set of all clause-sets is denoted by $\Cls$. For $k \in \NNZ$ we use $\Pcls{k} := \set{F \in \Cls \mb \fa\, C \in F : \abs{C} \le k}$ for the set of clause-sets where all clauses have length at most $k$.

A special clause is the empty clause $\bot := \es \in \Cl$, and a special clause-set is the empty clause-set $\top := \es \in \Cls$. By $\lit(F) := \bc F \cup \ol{\bc F}$ we denote the set of literals occurring at least in one polarity in $F$.

We use $\var: \Lit \ra \Va$ for the underlying variable of a literal, $\var(C) := \set{\var(x) : x \in C} \subset \Va$ for the set of variables in a clause, and $\var(F) := \bc_{C \in F} \var(C)$ for the set of variables in a clause-set. So $\lit(F) = \var(F) \cup \ol{\var(F)}$. The number of variables in a clause-set is $n(F) := \abs{\var(F)} \in \NNZ$, the number of clauses is $c(F) := \abs{F} \in \NNZ$, and the number of literal occurrences is $\ell(F) := \sum_{C \in F} \abs{C} \in \NNZ$.

A \emph{full clause-set} is a clause-set $F$ such that each clause contains all variables, that is, for all $C \in F$ we have $\var(C) = \var(F)$. The set of Horn clause-sets is $\Ho \subset \Cls$, where every clause contains at most one positive literal, while $\Ho^+ \subset \Ho$ is the set of pure Horn clause-sets, where every clause contains exactly one positive literal. $\Ho \subset \Rho \subset \Cls$ is the set of renamable (``hidden'') Horn clause-sets, which by flipping signs can be turned into a Horn clause-set.

A partial assignment $\vp: V \ra \set{0,1}$ maps a finite $V \subset \Va$ to truth-values, the set of all partial assignments is $\Pass$. A special partial assignment is the empty partial assignment $\epa := \es \in \Pass$. We can construct partial assignments via $\pab{v_1 \ra \ve_1, \dots, v_n \ra \ve_n} \in \Pass$ for $v_i \in \Va$ and $\ve_i \in \set{0,1}$ (which must be consistent). We use $\var(\vp) := V = \dom(\vp)$ for the variables in the domain of $\vp$, and by $\Tass(V)$ we denote the set of all ``total assignments'' for $V$, that is, the $\vp \in \Pass$ with $\var(\vp) = V$. And $n(\vp) := \abs{\var(\vp)} \in \NNZ$ is the number of variables assigned by $\vp$.

For a partial assignment $\vp \in \Pass$ and a clause-set $F \in \Cls$ the application of $\vp$ to $F$ is denoted by $\vp * F \in \Cls$, which results from $F$ by removing all satisfied clauses (containing at least one satisfied literal), and removing all falsified literals from the remaining clauses. A class $\mc{C} \sse \Cls$ of clause-sets is \emph{stable under (application of) partial assignments} if for all $F \in \mc{C}$ and $\vp \in \Pass$ holds $\vp * F \in \mc{C}$.

A clause-set $F$ is satisfiable (i.e., $F \in \Sat \subset \Cls$) if there exists a partial assignment $\vp$ with $\vp * F = \top$, otherwise $F$ is unsatisfiable (i.e., $F \in \Usat := \Cls \sm \Sat$). For a clause $C$ the partial assignment $\bmm{\vp_C} \in \Pass$ is defined as $\vp_C := \pab{x \ra 0 : x \in C}$, that is, it sets precisely the literals of $C$ to $0$ (and leaves all other variables unassigned). For example $\vp_{\bot} = \epa$ and $\vp_{\set{x}} = \pao x0$.

Two clauses $C, D \in \Cl$ are resolvable if they clash in exactly one literal $x$, that is, $C \cap \ol{D} = x$, in which case their resolvent is $(C \cup D) \sm \set{x,\ol{x}}$ (with resolution literal $x$). A resolution tree is a binary tree formed by the resolution operation. We write \bmm{T : F \vdash C} if $T$ is a resolution tree with axioms (the clauses at the leaves) all in $F$ and with derived clause (at the root) $C$. By \bmm{\comptr(F)} for unsatisfiable $F$ the minimum number of leaves in a tree-resolution-refutation $T : F \vdash \bot$ is denoted.

A boolean function $f$ is a map $f: \Tass(V) \ra \set{0,1}$ for some finite $V =: \var(f)$; we can also use $f(\vp) \in \set{0,1}$ for $\vp \in \Pass$ with $\var(f) \sse \var(\vp)$, in which case $\vp$ is restricted to $\var(f)$. Special boolean functions are $0^V$ and $1^V$ for the constant-0 resp.\ constant-1 functions with domain $V$. We write $f \models g$ for boolean functions $f, g$ if for all partial assignments $\vp$ with $\var(\vp) \supseteq \var(f) \cup \var(g)$ we have $f(\vp) = 1 \Ra g(\vp) = 1$. Equivalence of boolean functions $f, g$ means $f \models g$ and $g \models f$ (so all $0^V$ are equivalent, and all $1^V$ are equivalent).

The interpretation of clauses $C$ and clause-sets $F$ as boolean functions is explicitly denoted by $\Cnf(C)$ and $\Cnf(F)$, using the CNF-interpretation (a clause as a disjunction of literals, a clause-set as a conjunction of clauses), and happens in this report typically implicitly.

For a boolean function $f$ the set of prime implicates is denoted by $\primec_0(f)$, the set of all clauses $C$ with $f \models C$ while for $C' \subset C$ holds $f \not\models C'$. (The ``$0$'' in $\primec_0(f)$ resp.\ $\primec_0(F)$ in the set of prime implicates of a boolean function or a clause-set (interpreted as CNF) shall remind at ``false'' or ``unsatisfiable'', since CNF have ``falsity'' at the core.) So a boolean function $f$ is equivalent to $\primec_0(f)$, that is, more explicitly, to $\Cnf(\primec_0(f))$. As it is well-known, by considering any clause-set $F$ equivalent to $f$ and computing the resolution-closure of $F$, followed by subsumption-elimination, we obtain precisely $\primec_0(f)$.

We denote by $\Cnf(f)$ the ``distinguished canonical normal form'', or the set of ``minterms of $f$'', that is, the set of clauses $C \in \Cl$ with $\var(C) = \var(f)$ and $f \models C$ (that is, $f \models \Cnf(C)$). Dually, by $\Dnf(f)$ we denote the set of clauses $C \in \Cl$ with $\var(C) = \var(f)$ and $\Dnf(C) \models f$ (the ``maxterms of $f$''; note that for us a clause is a combinatorial object, and the logical interpretation has to be added). In the DNF-interpretation a clause is the conjunction of its literals, and a clause-set is the disjunction of its clauses.

Finally, by $\rk_1: \Cls \ra \Cls$ unit-clause propagation is denoted, that is applying $F \leadsto \pao x1 * F$ as long as there are unit-clauses $\set{x} \in F$, and reducing $F \leadsto \set{\bot}$ in case of $\bot \in F$. In Definition \ref{def:rk} the general $\rk_k: \Cls \ra \Cls$ is defined.

\section{The SLUR class and extensions}
\label{sec:SLURh}

The SLUR-algorithm and the class $\Slur \subset \Cls$ have been introduced in \cite{SAFS95}. The SLUR-algorithm for input $F \in \Cls$ is an incomplete polynomial-time SAT algorithm, which either returns ``SAT'', ``UNSAT'' (in both cases correctly) or gives up. This algorithm is non-deterministic, and $\Slur$ is the class of clause-sets where it never gives up (and thus SAT-decision for $F \in \Slur$ can be done in polynomial time). Due to an observation attributed to Truemper in \cite{Fr96}, the SLUR-algorithm can be implemented such that it runs in linear time. Decision of membership, that is whether $F \in \Slur$ holds, by definition is in coNP, but only in \cite{CepekKuceraVlcek2012SLUR} it was finally shown that this decision problem is coNP-complete.

The original motivation was that $\Slur$ contains several other classes, including renamable Horn, extended Horn, hidden extended Horn, simple extended Horn and CC-balanced clause-sets, where for each class it was known that the SAT problem is solvable in polynomial time, but with in some cases rather complicated proofs, while it is trivial to see that the SLUR-algorithm runs in polynomial time. In \cite{Fr96,FrGe98} probabilistic properties of $\Slur$ have been investigated.\footnote{At this point a popular misunderstanding should be avoided: The well-known dichotomy result of Schaefer (see Subsection \ref{sec:Schaeffer}) states that under certain conditions there are precisely six classes of problem instances with polytime SAT solving (unless P=NP). However this has no bearing on the classes considered here, since they do not fall within the restricted framework of Schaefer's theorem.}

In this section we first give a semantic definition of $\Slur$ in Subsection \ref{sec:SLURalone}. In a nutshell, $\Slur$ is the class of clause-sets where either UCP (unit-clause propagation aka $\rk_1$) creates the empty clause, or where otherwise iteratively making assignments followed by UCP will always yield a satisfying assignment, given that these transitions do not obviously create unsatisfiable results, i.e., do not create the empty clause. In order to understand this definition (and its various extensions) clearly, we present a precise mathematical (non-algorithmic) definition, based on the transition relation $F \rslur F'$ (Definition \ref{def:slur2}), which represents one non-deterministic step of the SLUR algorithm: If $\rk_1$ on input $F \in \Cls$ does not determine unsatisfiability (in which case we have $F \in \Slur$), then $F \in \Slur$ iff $\top$ can be reached by this transition relation, while everything else reachable from $F$ is not an end-point of this transition relation.

 In \cite{CepekKuceraVlcek2012SLUR,BalyoGurskyKuceraVlcek2012SLURHier} recently three approaches towards generalising $\Slur$ have been considered, and we discuss them in Subsection \ref{sec:sluralt}. Our generalisation, called $\Slur_k$, which we see as the natural completion of these approaches, will be presented in Section \ref{sec:slurhier}.

\subsection{SLUR}
\label{sec:SLURalone}

The SLUR-algorithm (``Single Lookahead Unit Resolution'') from \cite{SAFS95} is described for input $F \in \Cls$ as follows:
\begin{enumerate}
\item First run UCP, that is, reduce $F \leadsto \rk_1(F)$.
\item If now $\bot \in F$ then we determined $F$ unsatisfiable.
\item If not, then the algorithm guesses a satisfying assignment for $F$, by repeated transitions $F \rslur F'$, where $F'$ is obtained by assigning one variable and then performing UCP, i.e., $F' = \rk_1(\pao x1 * F)$ for some literal $x$.
\item The ``lookahead'' means that a transition with $F' = \set{\bot}$ is avoided.
\item The algorithm might find a satisfying assignment in this way, or it gets stuck, that is, for the chosen literal both assignments $x \ra 1$ and $\ol{x} \ra 1$ yield $\set{\bot}$, in which case it ``gives up''.
\end{enumerate}
The SLUR class is defined as the class of clause-sets where this algorithm never gives up. The precise details are as follows. First we define the underlying transition relation (one non-failing transition from $F$ to $F'$):
\begin{defi}\label{def:slur1}
  For clause-sets $F, F' \in \Cls$ the relation \bmm{F \rslur F'} holds if there is $x \in \lit(F)$ such that $F' = \rk_1(\pao x1 * F)$ and $F' \not= \set{\bot}$. The transitive-reflexive closure is denoted by \bmm{F \rslurs F'}.
\end{defi}
\begin{examp}\label{exp:rslur}
  Considering when we have $F \rslurs F'$ and when not:
  \begin{enumerate}
  \item $F \rslurs \top$ iff $F \in \Sat$.
  \item $\set{C} \rslur \top$ precisely for all clauses $C \not= \bot$.
  \item $\set{\set{x,y},\set{x,\ol{y}}} \rslur \top$.
  \item $\set{\set{\ol{x},y},\set{\ol{y},z}} \rslur \top$ (due to e.g.\ $\rk_1(\pao x1 * \set{\set{\ol{x},y},\set{\ol{y},z}}) = \top$).
  \item $F \rslur F'$ does not hold if there is no literal to set, or if $\rk_1$ detects unsatisfiability of $F'$. That is, there are \textbf{no} clause-sets $F, F'$ such that any of the following hold:
    \begin{enumerate}
    \item $\top \rslur F$.
    \item $\set{\bot} \rslur F$.
    \item $F \rslur F$.
    \item $F \rslur F'$ where $\rk_1(F') = \set{\bot}$.
    \end{enumerate}
  \end{enumerate}
\end{examp}

Via the transition-relation $F \rslur F'$ we can now easily define the class $\Slur$, which will find a natural generalisation in Definition \ref{def:kslur} to $\Slur_k$ for $k \in \NNZ$ (where $\Slur = \Slur_1$):
\begin{defi}\label{def:slur2}
  The set of all fully reduced clause-sets reachable from $F \in \Cls$ is denoted by
  \begin{displaymath}
    \bmm{\slur(F)} := \set{F' \in \Cls \mb F \rslurs F' \und \neg \ex\, F'' \in \Cls : F' \rslur F''}.
  \end{displaymath}
  Finally the class of all clause-sets which are either identified by UCP to be unsatisfiable, or where by SLUR-reduction always a satisfying assignment is found, is denoted by $\bmm{\Slur} := \set{F \in \Cls : \rk_1(F) \not= \set{\bot} \Ra \slur(F) = \set{\top}}$.
\end{defi}
We could define $\rslur$ as $F \rslur \pao x1 * F$ iff $\rk_1(\pao x1 * F) \not= \bot$, and this would yield the same class $\Slur$ but a different transition relation (one would not be forced to immediately make forced assignments).

\begin{examp}\label{exp:slur}
  Computing $\slur(F)$ for clause-sets $F$:
  \begin{enumerate}
  \item $\slur(F) \not= \es$ (in the ``worst'' case we have $F \in \slur(F)$).
  \item $\slur(\set{\bot}) = \set{\set{\bot}}$.
  \item $\slur(\top) = \set{\top}$.
  \item $\slur(\set{C}) = \set{\top}$ iff $C \not= \bot$.
  \item If $\rk_1(F) = \top$ then $\slur(F) = \set{\top}$.
  \item $\slur(\set{\set{x,y},\set{x,\ol{y}}}) = \set{\top}$.
  \item $\slur(\set{\set{\ol{x},y},\set{\ol{y},z}}) = \set{\top}$.
  \item For $F := \set{\set{x,y},\set{x,\ol{y}},\set{\ol{x},y},\set{\ol{x},\ol{y}}}$ we have $\slur(F) = \set{F}$.
  \item For $F' := \set{\set{z,x,y},\set{z,x,\ol{y}},\set{z,\ol{x},y},\set{z,\ol{x},\ol{y}}}$ we have $\top, F \in \slur(F')$.
  \end{enumerate}
\end{examp}

\subsection{Previous approaches for SLUR hierarchies}
\label{sec:sluralt}

In \cite{CepekKuceraVlcek2012SLUR,BalyoGurskyKuceraVlcek2012SLURHier} three hierarchies $\Altsluri{k}, \Altslurstari{k}$ ($k \in \NN$) and $\Canoni{k}$ ($k \in \NNZ$) have been introduced. In Section 4 of \citep{BalyoGurskyKuceraVlcek2012SLURHier} it is shown that $\Altsluri{k} \subset \Altslurstari{k}$ for all $k \in \NN$ and so we restrict our attention to $\Altslurstari{k}$ and $\Canoni{k}$.

$\Canoni{k}$ is defined to be the set of clause-sets $F$ such that every $C \in \primec_0(F)$ can be derived from $F$ by a resolution tree of height at most $k$. Note that basically by definition (using stability of resolution proofs under application of partial assignments) we get that each $\Canoni{k}$ is stable under application of partial assignments and under variable-disjoint union.

The $\Altslurstari{k}$ hierarchy is derived in \citep{BalyoGurskyKuceraVlcek2012SLURHier} from the $\Slur$ class by extending the reduction $r_1$. We provide an alternative formalisation here, in the same manner as in Section \ref{sec:SLURalone}. The main question is the transition relation $F \leadsto F'$. The $\Altslurstari{k}$-hierarchy provides stronger and stronger witnesses that $F'$ might be satisfiable, by longer and longer assignments (making ``$k$ decisions'') not yielding the empty clause:
\begin{defi}\label{def:kdecisions}
  That partial assignment $\vp \in \Pass$ \textbf{makes $k$ decisions} for some $k \in \NNZ$ w.r.t.\ $F \in \Cls$ is defined recursively as follows: For $k = 0$ this relation holds if $\vp * F = \rk_1(F)$, while for $k > 0$ this relation holds if either there is $k' < k$ such that $\vp$ makes $k'$ decision w.r.t.\ $F$ and $\vp * F = \top$, or there exists $x \in \lit(F)$ and a partial assignment $\vp'$ making $k-1$ decision for $\rk_1(\pao x1 * F)$, and where $\vp * F = \vp' * \rk_1(\pao x1 * F)$.

  Now \bmm{F \rkslurstar{k} F'} for $k \ge 1$ by definition holds if there is a partial assignment $\vp$ making $k$ decision w.r.t.\ $F$ with $F' = \vp * F$, where $F' \not= \set{\bot}$. The reflexive-transitive closure is $\rkslursstar{k}$. 
\end{defi}
Finally we can define the hierarchy:
\begin{eqnarray*}
  \bmm{\slurstar(k)(F)} &:=& \set{F' \in \Cls \mb F \rkslursstar{k} F' \und \neg \ex\, F'' : F' \rkslurstar{k} F''}\\
  \bmm{\Altslurstari{k}} &:=& \set{F \in \Cls : \slurstar(k)(F) \ne \set{F} \Ra \slurstar(k)(F) = \set{\top}}.
\end{eqnarray*}
The unsatisfiable elements of $\Altslurstari{k}$ are those $F \not= \top$ with $\slurstar(k)(F) = \set{F}$. By definition each $\Altslurstari{k}$ is stable under application of partial assignments, but not stable under variable-disjoint union, since the number of decision variables is bounded by $k$ (in Lemma \ref{lem:stabuc} we will see that our hierarchy is stable under variable-disjoint union, which is natural since it strengthens the $\Canoni{k}$-hierarchy).

\begin{examp}\label{exp:altslur}
  Some examples for $\Canoni{k}$ and $\Altslurstari{k}$ ($k \in \NN$):
  \begin{enumerate}
  \item Consider the unsatisfiable clause-set $F := \set{\set{x,y},\set{x,\ol{y}},\set{\ol{x},y},\set{\ol{x},\ol{y}}}$.
    \begin{enumerate}
    \item $F \not\in \Slur$ because $F$ is unsatisfiable but $\rk_1(F) \not= \set{\bot}$.
    \item $F \in \Altslurstari{1}$ because $\rk_1(\pao {x'}1 * F) = \set{\bot}$ for all $x' \in \lit(F)$ and so $\slurstar(1)(F) = \set{F}$.
    \item This establishes $\Slur \subset \Altslurstari{1}$.
    \item $F \in \Canoni{2} \sm \Canoni{1}$ because actually all tree-resolution refutations of $F$ are full binary trees of height $2$.
    \end{enumerate}
  \item Consider the satisfiable clause-set $F' := \set{ \set{x_1,\dots,x_k} \cup C \mb C \in F }$.
    \begin{enumerate}
    \item $F' \not\in \Altslurstari{k}$ because $F' \rkslursstar{k} F$, where $F$ is unsatisfiable and thus $\neg (F \rkslursstar{k} \top)$, whence $\slurstar(k)(F') \not= \set{\top}$.
    \item $F' \in \Altslurstari{k+1}$ because we have $r_1(\vp * F') \in \set{\top, \set{\bot}}$ for all partial assignments $\vp$ of length $k+1$ on variables of $F'$ hence $\slurstar(k)(F_1) = \set{\top}$.
    \item $F' \in \Canoni{2}$ because the only prime implicate is $\set{x_1,\dots,x_k}$ and actually all its tree-resolution proofs are full binary trees of height $2$.
    \end{enumerate}
  \end{enumerate}
\end{examp}

\section{Generalised unit-clause propagation}
\label{sec:genucp}

In this section we review the approximations of forced assignments as computed by the hierarchy of reductions $\rk_k: \Cls \ra \Cls$ from \citep{Ku99b,Ku00g} for $k \in \NNZ$. First we introduce the semantical notion of forced literals/assignments in Subsection \ref{sec:forcedlitass} together with the limit-reduction $\rki: \Cls \ra \Cls$, which eliminates \emph{all} forced assignments. In Subsection \ref{sec:hierred} then the $\rk_k$-reductions themselves (eliminating some forced assignments) are defined and basic properties discussed. In Subsection \ref{sec:Hardnesstrees} finally we introduce generalised (nested) input resolution and its main parameter, the ``Horton-Strahler number'' of the corresponding resolution tree, generalising the well-known refutational equivalence between unit resolution and input resolution, and providing the proof-theoretic background.

For further discussions of these reductions, in the context of SAT decision and in their relations to various consistency and width-related notions, see \citep{Ku99b,Ku00g} and Section 3 in \citep{Kullmann2007Uebersicht}. It seems to us that the $\rk_k$-reductions establish the SAT-counterpart to consistency-notions from the constraint literature (see \cite{Bessiere2006Propagation} for an overview). We have the following basic distinction between SAT and CSP: SAT has the extremely ``thin'' clauses, enabling the global point of view (``no (or flat) hierarchies''), while CSP has ``fat'' constraints, the ``lumping together'' of clauses. In the SAT world, the $\rk_k$-reductions approximate global consistency via approaching all assignments of $\rki$, while in the CSP world, consistency means making the constraints stronger and stronger (lumping more and more clauses together), until only one constraint is left. Thus the (stronger) consistency-notions of CSP are more related to width-restricted resolution, while, as shown in \citep{Ku99b,Ku00g}, the $\rk_k$-reductions are much weaker (each only using linear space). Making a clause-set $F$ ``consistent'' in the SAT world thus means (to us) to find a ``representation'' $F'$ of $F$ (see Subsection \ref{sec:outlinegoodrep} for some discussion on ``representations''), where via $\rk_k$ for some $k \in \NNZ$ we can derive ``everything'', which is embodied in its most elementary form in the $\Urefc_k$-hierarchy, that is, via the condition $F' \in \Urefc_k$ (Definition \ref{def:urefc}).

\subsection{Forced literals/assignments}
\label{sec:forcedlitass}

Fundamental is the notion of a ``forced literal'' of a boolean function resp.\ a clause-set\footnote{we prefer this logical (and common) terminology over ``backbone literal'', which is only used in a special context}, which are literals which must be set to true in order to satisfy the function resp.\ clause-set:
\begin{defi}\label{def:forcedliteral}
  A literal $x$ is \textbf{forced} for a boolean function $f$ if $f \models x$, and the set of forced literals for $f$ is $\bmm{\frl(f)} \sse \Lit$. A literal is forced for a clause-set $F$ if it is forced for $\Cnf(F)$, and we set $\frl(F) := \frl(\Cnf(F))$.
\end{defi}
Every literal is forced for every $0^V$. In fact a boolean function $f$ is constant zero iff $\frl(f) = \Lit$ iff there is a literal $x$ with $x, \ol{x} \in \frl(f)$. No literal is forced for any $1^V$ (i.e., $\frl(1^V) = \es$). We have for every boolean function $f$ that
\begin{displaymath}
  \frl(f) = \rbca{\Lit} \Dnf(f)
\end{displaymath}
(the index ``$\Lit$'' in the intersection is the ``universe'' of the sets considered in the intersection, which becomes the result if there are no sets to intersect, that is, if $f$ is unsatisfiable). More directly we can read off the forced literals from the prime clauses, namely $x$ is forced for $f$ iff $\primec_0(f) \cap \set{\bot,\set{x}} \not= \es$.

\begin{examp}\label{exp:forcedlits}
  Here are some basic determinations of $\frl(F)$:
  \begin{enumerate}
  \item $\frl(\set{\bot}) = \Lit$.
  \item $\frl(\top) = \es$.
  \item $\frl(\set{\set{x_1},\dots,\set{x_n}}) = \set{x_1,\dots,x_n}$.
  \item $\frl(\set{\set{x,\ol{y}},\set{\ol{x},y}}) = \es$.
  \item $\frl(\set{\set{x,y},\set{x,\ol{y}}}) = \set{x}$.
  \end{enumerate}
\end{examp}

If $x$ is a forced literal for $F$, then the \textbf{forced assignment} $\pao x1$ yields the clause-set $\pao x1 * F$ which is satisfiability-equivalent to $F$. We denote by $\bmm{\rki(F)} \in \Cls$ the result of applying all forced assignments to $F$. Note that $F$ is unsatisfiable iff $\rki(F) = \set{\bot}$ (while $F$ is uniquely satisfiable after discarding variables without influence iff $\rki(F) = \top$).

\subsection{A hierarchy of reductions}
\label{sec:hierred}

We now review the hierarchy $\rk_k: \Cls \ra \Cls$, $k \in \NNZ$, of reductions (\citep{Ku99b}), which achieves approximating $\rki$ by poly-time computable functions. The basic idea is that unit-clause propagation in a sense computes the most direct forced assignments (at ``level $k=1$''), and generalisations like failed-literal elimination (level $k=2$) find more forced assignments.
\begin{defi}[\citep{Ku99b}]\label{def:rk}
  The maps $\bmm{\rk_k}: \Cls \ra \Cls$ for $k \in \NNZ$ are defined as follows (for $F \in \Cls$):
  \begin{eqnarray*}
    \rk_0(F) & := &
    \begin{cases}
      \set{\bot} & \text{if } \bot \in F\\
      F & \text{otherwise}
    \end{cases}\\
    \rk_{k+1}(F) & := &
    \begin{cases}
      \rk_{k+1}(\pao x1 * F) & \text{if } \ex\, x \in \lit(F) : \rk_k(\pao x0 * F) = \set{\bot}\\
      F & \text{otherwise}
    \end{cases}.
  \end{eqnarray*}
\end{defi}
$\rk_1$ is unit-clause propagation, $\rk_2$ is (full) failed literal elimination. We call $\rk_k$ \textbf{generalised unit-clause-propagation of level $k$}. In \citep{Ku99b} one finds the following basic observations proven (for $k \in \NNZ$, $F \in \Cls$ and $\vp \in \Pass$):
\begin{itemize}
\item The map $\rk_k: \Cls \ra \Cls$ is well-defined (does not depend on the choices).
\item $\rk_k$ applies only forced assignments (and so $\rk_k(F)$ is satisfiability-equivalent to $F$).
\item $\rk_k(F)$ is computable in time $O(\ell(F) \cdot n(F)^{2(k-1)})$ and linear space.
\item $\rk_k(F) = \set{\bot}$ implies $\rk_k(\vp * F) = \set{\bot}$.
\item $\rk_k(\vp * \rk_k(F)) = \rk_k(\vp * F)$.
\end{itemize}
Quasi-automatisation of tree-resolution is achieved for inputs $F \in \Usat$ by applying $\rk_0(F), \rk_1(F), \dots$ until unsatisfiability has been achieved (\citep{Ku99b}). Also satisfiable instances are handled in \citep{Ku99b}, however in this paper we do not consider these algorithmical aspects.

Actually, a more general form was introduced in \citep{Ku99b}, namely $\rk_k^{\mc{U}}$ for some oracle $\mc{U}$ deciding unsatisfiability at level $0$. We believe that this generalisation is important for further progress (see Subsection \ref{sec:outlinerelhd}), however in this report we only consider the trivial oracle $\mc{U} = \set{F \in \Cls : \bot \in F}$, which (only) recognises unsatisfiability at level $0$ iff the empty clause occurs. A further generalisation to constraint-like systems (via an abstract, axiomatic approach) was achieved in \citep{Ku00g}, however in this initial study we do only consider boolean values and CNF-representations.

\begin{examp}\label{exp:rk}
  Computing some $\rk_k(F)$ (using literals $x_1,\dots,x_n,x,y$ with pairwise different underlying variables):
  \begin{enumerate}
  \item $\rk_k(\set{\bot}) = \set{\bot}$ for $k \ge 0$.
  \item $\rk_k(\top) = \top$ for $k \ge 0$.
  \item For $F := \set{\set{x_1},\dots,\set{x_n}}$: $\rk_0(F) = F$, $\rk_k(F) = \top$ for $k \ge 1$.
  \item For $F' := F \cup \set{\set{x,y}}$: $\rk_0(F') = F'$, $\rk_k(F') = \set{\set{x,y}}$ for $k \ge 1$ (note that $\set{\set{x,y}}$ has no forced assignments).
  \item For $F := \set{\set{x,y},\set{x,\ol{y}}}$: $\rk_k(F) = F$ for $k \le 1$, $\rk_k(F) = \top$ for $k \ge 2$.
  \item For $F := \set{\set{x,y},\set{x,\ol{y}}, \set{\ol{x},y},\set{\ol{x},\ol{y}}}$: $\rk_k(F) = F$ for $k \le 1$, $\rk_k(F) = \set{\bot}$ for $k \ge 2$.
  \end{enumerate}
\end{examp}
Via the reductions $\rk_k$ we can approximate the implication relation $F \models C$ as follows:
\begin{defi}[\citep{Ku99b,Ku00g}]\label{def:implick}
  For $k \in \NNZ$, clause-sets $F$ and clauses $C$ the relation \bmm{F \implk{k} C} holds if $\rk_k(\vp_C * F) = \set{\bot}$.
\end{defi}
As it is well-known, $F \implk{1} C$ iff some subclause of $C$ follows from $F$ via input resolution.

\begin{examp}\label{exp:implk}
  Consider $k \in \NNZ$ and literals $x,y,w$:
  \begin{enumerate}
  \item For all $k \ge 0$ and all clauses $C$ we have:
    \begin{enumerate}
    \item $F \implk{k} C$ if there is $D \in F$ with $D \sse C$ (note $\bot \in \vp_C * F$).
    \item $\set{\bot} \implk{k} C$ and $\top \not\implk{k} C$.
    \end{enumerate}
  \item $\set{\set{x,y},\set{x,\ol{y}}} \implk{k} \set{x}$ iff $k \ge 1$.
  \item For $F := \set{\set{\ol{x},y}, \set{\ol{y},z}}$ we have $F \implk{k} \set{\ol{x},z}$ iff $k \ge 1$.
  \item For $F := \set{\set{\ol{x},y,w}, \set{\ol{y},z,w},\set{\ol{x},y,\ol{w}}, \set{\ol{y},z,\ol{w}}}$ we have $F \implk{k} \set{\ol{x},z}$ iff $k \ge 2$ (note that $\pab{x \ra 1, z \ra 0} * F \in \Pcls{2}$).
  \end{enumerate}
\end{examp}

\subsection{Generalised input resolution}
\label{sec:Hardnesstrees}

In \citep{Ku99b}, Chapter 4, the \emph{levelled height} ``$h(T)$'' of branching trees $T$ has been introduced, which was further generalised in \citep{Ku00g}, Chapter 3 (to a general form of constraint satisfaction problems). It handles satisfiable as well as unsatisfiable clause-sets. In this report we will only use the unsatisfiable case. In this case the measure reduces to a well-known measure which only considers the structure of the tree. As discussed in Subsections 4.2, 4.3 of \citep{Ku99b}, this case, the levelled height of splitting trees for unsatisfiable clause-sets, appeared at many places in the literature. \cite{AnsoteguiBonetLevyManya2008Hardness} used the term ``Horton-Strahler number'' (sometimes also ``\href{http://en.wikipedia.org/wiki/Strahler_number}{Strahler number}''): it seems the oldest source (from 1945), however disconnected from its various (re-)inventions in computer science. As in \cite{AnsoteguiBonetLevyManya2008Hardness}, the Horton-Strahler number of the trivial tree is $0$.

\begin{defi}\label{def:hdtree}
  Consider a resolution tree $T$. The \textbf{Horton-Strahler number} $\bmm{\hts(T)} \in \NNZ$ is defined as $\hts(T) := 0$, if $T$ is trivial (consists only of one node), while otherwise we have two subtrees $T_1, T_2$, and we set $\hts(T) := \max(\hts(T_1),\hts(T_2))$ if $\hts(T_1) \not= \hts(T_2)$, while in case of $\hts(T_1) = \hts(T_2)$ we set $\hts(T) := \max(\hts(T_1),\hts(T_2)) + 1$.
\end{defi}
See Sections 4.2, 4.3 in \citep{Ku99b} for various characterisations of $\hts(T)$.

\begin{examp}\label{exp:hortonstrahler}
  Examples of trees with their Horton-Strahler numbers. We denote by $T_1$ and $T_2$ in each example the left and right sub-trees of the root.
  \begin{center}
    \begin{tabular}{|>{\centering\arraybackslash}m{10em}|>{\centering\arraybackslash}m{3em}|>{\centering\arraybackslash}m{7em}|}
      \hline
      \textbf{Tree $T$} & \bmm{\hts(T)} & \textbf{Explanation} \\\hline\hline
      $\displaystyle
      \xygraph{
        []{\cdot} ( )
      }$ & 0 & trivial tree \\ \hline
      $\displaystyle
      \xygraph{
        !{0;/r3ex/:}
        []{\cdot} (
          - [dl]{\cdot} (),
          - [dr]{\cdot} ()
        )
      }$ & 1 & $\hts(T_1) = 0$, $\hts(T_2) = 0$. \\ \hline
      $\xygraph{
        !{0;/r3ex/:}
        []{\cdot} (
        - [dl]{\cdot} (),
        - [dr]{\cdot} (
          -[dl]{\cdot} (),
          -[dr]{\cdot} ()
        )
        )
      }$ & 1 & $\hts(T_1) = 0$, $\hts(T_2) = 1$.  \\ \hline
      $\xygraph{
        !{0;/r3ex/:}
        []{\cdot} (
          - [dl]{\cdot} (),
          - [dr]{\cdot} (
            -[dl]{\cdot} (),
            - [dr]{\cdot} (
              -[dl]{\cdot} (),
              -[dr]{\cdot} ()
            )
          )
        )
      }$ & 1 & $\hts(T_1) = 0$, $\hts(T_2) = 1$.  \\ \hline
      $\xygraph{
        !{0;/r3ex/:}
        []{\cdot} (
        - [dll]{\cdot} (
          -[dl]{\cdot} (),
          -[dr]{\cdot} ()
        ),
        - [drr]{\cdot} (
          -[dl]{\cdot} (),
          -[dr]{\cdot} ()
        )
        )
      }$ & 2 & $\hts(T_1) = 1$, $\hts(T_2) = 1$.  \\ \hline
      $\xygraph{
        !{0;/r3ex/:}
        []{\cdot} (
        - [dll]{\cdot} (
          -[dl]{\cdot} (),
          -[dr]{\cdot} ()
        ),
        - [drr]{\cdot} (
          -[dll]{\cdot} (
            -[dl]{\cdot} (),
            -[dr]{\cdot} ()
          ),
          -[dr]{\cdot} (
            -[dl]{\cdot} (),
            -[dr]{\cdot} ()
          )
        )
        )
      }$ & 2 & $\hts(T_1) = 1$, $\hts(T_2) = 2$.  \\ \hline
    \end{tabular}
  \end{center}
\end{examp}

In \citep{Ku99b}, Section 7 (generalised in \citep{Ku00g}, Section 5), \emph{generalised input resolution} was introduced. We use the notation ``$\vdash_k$'' for it:
\begin{defi}[\citep{Ku99b,Ku00g}]\label{def:geninpres}
  For a clause-set $F \in \Cls$ and a clause $C \in \Cl$ the relation \bmm{F \vdash_k C} ($C$ can be derived from $F$ by \textbf{$k$-times nested input resolution}) holds if there exists a resolution tree $T$ and $C' \sse C$ with $T : F \vdash C'$ and $\hts(T) \le k$.
\end{defi}
By parts 1 and 2 of Theorem 7.5 in \citep{Ku99b}, generalised in Corollary 5.12 in \citep{Ku00g}:
\begin{lem}[\citep{Ku99b,Ku00g}]\label{lem:equivrkinpres}
  For clause-sets $F$, clauses $C$ and $k \in \NNZ$ we have $F \implk{k} C$ if and only if $F \vdash_k C$.
\end{lem}

\section{Hardness}
\label{sec:Hardness}

This section is devoted to the discussion of $\hardness: \Cls \ra \NNZ$. It is the central concept of the paper, from which the hierarchy $\Urefc_k$ is derived (Definition \ref{def:urefc}). The basic idea is to start with some measurement $h: \Usat \ra \NNZ$ of ``the complexity'' of unsatisfiable $F$. This measure is extended to arbitrary $F \in \Cls$ by maximising over all ``sub-instances'' of $F$, that is, over all unsatisfiable $\vp * F$ for (arbitrary) partial assignments $\vp$. A first guess for $h: \Usat \ra \NNZ$ is to take something like the logarithm of the tree-resolution complexity of $F$. However this measure is too fine-grained, and doesn't yield a hierarchy like $\Urefc_k$, where each level brings a qualitative enhancement. Another approach is algorithmical, measuring how far $F$ is from being refutable by unit-clause propagation. As shown in \citep{Ku99b,Ku00g}, actually these two lines of thought can be brought together by the hardness measure $\hardness: \Usat \ra \NNZ$. Why only tree-resolution, and not dag-resolution (i.e., full resolution)? The tree-resolution approach is the natural starting point, and what is easy for tree-resolution is also easy for dag-resolution. Our basic approach towards the more complicated handling of dag-resolution is shown in Subsection \ref{sec:outlinewhd}.

The outline of this section is as follows. $\hardness(F)$ is defined and discussed for unsatisfiable $F$ in Subsection \ref{sec:harducls}. The general case (arbitrary $F$) is handled in Subsection \ref{sec:hardcls} by reduction to the unsatisfiable cases within $F$ (as produced by applying partial assignments). The central result of this section can be seen in Theorem \ref{thm:characuckir}, which shows that $F \in \Urefc_k$ (i.e., $\hardness(F) \le k$) is equivalent to the condition that all prime implicates of $F$ can be derived by some resolution tree with a Horton-Strahler number at most $k$. In this way some form of geometric intuition is gained, and a machinery becomes available. The first applications are given by the various lemmas in Section \ref{sec:fundprop} for determining hardness under various circumstances.

We remark that, when considering only unsatisfiable clause-sets $F$, in \citep{Ku99b,Ku00g} actually a general concept of ``hardness'' was introduced, parameterised by an oracle $\mc{U} \sse \Usat$ for (``easy'') detection of special cases of unsatisfiability. In this report only $\mc{U} = \set{F \in \Cls : \bot \in F}$ is used, but we expect the general theory to become important in the future. See Subsection \ref{sec:outlinerelhd} for some further discussions.

\subsection{Hardness of unsatisfiable clause-sets}
\label{sec:harducls}

In \citep{Ku99b} the following hardness parameter was introduced and investigated (further generalised in \citep{Ku00g}):
\begin{defi}[\citep{Ku99b,Ku00g}]\label{def:hardnessucls}
  The \textbf{hardness} $\hardness(F)$ of an unsatisfiable $F \in \Cls$ is the minimal $k \in \NNZ$ such that $\rk_k(F) = \set{\bot}$.
\end{defi}
As shown in \citep{Ku99b}, $\hardness(F)+1$ is precisely the clause-space complexity of $F$ regarding tree-resolution (see \cite{Nordstrom2012PebbleProofTimeSpaceSurvey} for a recent overview on space complexity of resolution). In \citep{Ku99b,Ku00g} the notation ``$\mr{h}(F)$'' was used (resp., more generally, ``$\mr{h}_{\mc{U},\mc{S}}(F)$'', using oracles for unsatisfiability and satisfiability detection), which seems now to us too unspecific. From \cite{He74} we gain the insight that for $F \in \Usat$ holds $\hardness(F) \le 1$ iff there exists $F' \sse F$ which is an unsatisfiable renamable Horn clause-set (i.e., $F' \in \Rho \cap \Usat$). By Theorem 7.8 (and Corollary 7.9) in \citep{Ku99b} (or, more generally, Theorem 5.14 in \citep{Ku00g}) we have for $F \in \Usat$:
  \begin{displaymath}
    2^{\hardness(F)} \le \comptr(F) \le (n(F)+1)^{\hardness(F)}.
  \end{displaymath}

\begin{examp}\label{exp:harducls}
  Some basic determinations of $\hardness(F)$ for unsatisfiable $F$:
  \begin{enumerate}
  \item $\hardness(F) = 0$ iff $\bot \in F$.
  \item $\hardness(\set{\set{x},\set{\ol{x}}}) = 1$.
  \item $\hardness(\set{\set{x},\set{\ol{x},y}, \set{\ol{y},z}, \set{\ol{z}}}) = 1$.
  \item $\hardness(\set{\set{x,y},\set{x,\ol{y}},\set{\ol{x},y},\set{\ol{x},\ol{y}}}) = 2$.
  \item $\hardness(\set{\set{x,\ol{y}},\set{\ol{x},y},\set{y,\ol{z}},\set{\ol{y},z},\set{x,y,z},\set{\ol{x},\ol{y},\ol{z}}}) = 2$.
  \end{enumerate}
\end{examp}

By Lemma \ref{lem:equivrkinpres} we get:
\begin{lem}[\citep{Ku99b,Ku00g}]\label{lem:charachdhs}
  For an unsatisfiable clause-set $F$ and $k \in \NNZ$ we have $\hardness(F) \le k$ iff $F \implk{k} \bot$ iff $F \vdash_k \bot$.
\end{lem}
By applying partial assignments we can reach all hardness-levels in a clause-set, as the following lemma shows.
\begin{lem}\label{lem:hardnessgrad}
  For an unsatisfiable clause-set $F$ and every $0 \le k \le \hardness(F)$ there exists a partial assignment $\vp$ with $n(\vp) = k$ and $\hardness(\vp * F) = \hardness(F)-k$.
\end{lem}
\begin{prf}
We proceed by induction on $n(F)$. As $k \le \hardness(F) \le n(F)$, for the base case we consider $n(F) = k$. If $n(F) = k$ then all $\vp$ with $n(\vp) = k$ have $\hardness(\vp * F) = \hardness(\set{\bot}) = 0 = \hardness(F) - k$. For $n(F) > k$, we make a case distinction on the value of $k$. If $k = 0$ then choose $\vp = \epa$. If $k = 1$ then:
\begin{enumerate}
\item Assume for the sake of contradiction that there is no $x \in \lit(F)$ such that $\hardness(\pao{x}{1} * F) = \hardness(F) - 1$; otherwise we are done.
\item If for all $x \in \lit(F)$ we had $\hardness(\pao{x}{1} * F) \le \hardness(F) - 2$ then by Definition \ref{def:hardnessucls} we would have $\hardness(F) \le k - 1$, a contradiction.
\item Therefore there must exist an $x \in \lit(F)$ such that
  \begin{displaymath}
    \hardness(F) = \hardness(\pao{x}{1} * F) > \hardness(\pao{x}{0} * F) + 1.
  \end{displaymath}
\item By induction hypothesis we have a partial assignment $\vp$ with $n(\vp) = 1$ such that $\hardness(\vp * (\pao{x}{1} * F)) = \hardness(F) - 1$.
\item Application of partial assignments doesn't increase hardness (Lemma 3.11 of \citep{Ku99b}) and so we have
  \begin{displaymath}
    \hardness(\vp * F) \ge \hardness(\pao{x}{1} * (\vp * F)) = \hardness(F) - 1.
  \end{displaymath}
\item By our choice of $x$ we have
  \begin{eqnarray*}
    \hardness(\pao{x}{1} * (\vp * F)) & = & \hardness(F) - 1 \\
    \hardness(\pao{x}{0} * (\vp * F)) & \le & \hardness(F) - 2,
  \end{eqnarray*}
  therefore by Definition \ref{def:hardnessucls} we have $\hardness(\vp * F) \le \hardness(F) - 1$.
\item Thus we have that $\hardness(\vp * F) = \hardness(F) - 1$.
\end{enumerate}
Finally, for $k > 1$, we apply induction using the $k = 1$ case; once we can reduce by $1$ we can reduce by $k$. \Qed
\end{prf}

\subsection{Hardness of arbitrary clause-sets}
\label{sec:hardcls}

The hardness $\hardness(F)$ of arbitrary clause-sets can now be defined as the maximum hardness over all unsatisfiable instances obtained by partial assignments.
\begin{defi}\label{def:hardnesscls}
  The \textbf{hardness} $\bmm{\hardness(F)} \in \NNZ$ for $F \in \Cls$ is the minimal $k \in \NNZ$ such that for all clauses $C$ with $F \models C$ we have $F \implk{k} C$ (recall Definition \ref{def:implick}; by Lemma \ref{lem:equivrkinpres} this is equivalent to $F \vdash_k C$).
\end{defi}
In other words, if $F \not= \top$ then $\hardness(F)$ is the maximum of $\hardness(\vp * F)$ for partial assignments $\vp$ such that $\vp * F \in \Usat$. To our knowledge, the measure $\hardness(F)$ for satisfiable $F$ was mentioned the first time in the literature in \cite{AnsoteguiBonetLevyManya2008Hardness}, Definition 8 (the only result there concerning this measure is Lemma 9, relating it to another hardness-alternative for satisfiable $F$). Note that one can restrict attention in Definition \ref{def:hardnesscls} to $C \in \primec_0(F)$. Hardness $0$ means that all prime clauses are there, i.e., $\hardness(F) = 0$ iff $\primec_0(F) \sse F$. Especially $\hardness(\top) = 0$.

Lemma \ref{lem:hardnessgrad}, stating that $\hardness(\vp * F)$ takes exactly the values from $0$ to $\hardness(F)$, extends by definition to satisfiable $F \in \Cls$, when adding to the size of the partial assignment $\vp$ the minimum size of a partial assignment $\psi$ with $\psi * F \in \Usat$ and $\hardness(\psi * F) = \hardness(F)$.
\begin{defi}\label{def:urefc}
  For $k \in \NNZ$ let $\bmm{\Urefc_k} := \set{F \in \Cls : \hardness(F) \le k}$ (the class of \textbf{unit-refutation complete clause-sets of level $k$}).
\end{defi}
The class $\Urefc_1$ has been introduced in \cite{Val1994UnitResolutionComplete} for knowledge compilation. Various (resolution-based) algorithms computing for clause-sets $F$ some equivalent set $F' \in \Urefc_1$ of prime implicates are discussed there. Based on the results from \citep{Ku99b,Ku00g}, we can now give a powerful proof-theoretic characterisation for all classes $\Urefc_k$:
\begin{thm}\label{thm:characuckir}
  For $k \in \NNZ$ and $F \in \Cls$ we have
  \begin{displaymath}
    F \in \Urefc_k \iff \fa\, C \in \primec_0(F) : F \vdash_k C.
  \end{displaymath}
  Thus if every $C \in \primec_0(F)$ has a tree-resolution refutation using at most $2^{k+1}-1$ leaves (i.e., $\comptr(\vp_C * F) < 2^{k+1}$), then $\hardness(F) \le k$.
\end{thm}
\begin{prf}
The equivalence $F \in \Urefc_k \Lra \fa\, C \in \primec_0(F) : F \vdash_k C$ follows from Lemma \ref{lem:equivrkinpres}. And if $\hardness(F) > k$, then there is $C \in \primec_0(F)$ with $F \not\vdash_k C$, and then every tree-resolution derivation of $C$ from $F$ needs at least $2^{k+1}$ leaves due to $2^{\hardness(\vp_C * F)} \le \comptr(\vp_C * F)$ (as stated before). \Qed
\end{prf}

\begin{examp}\label{exp:hardcls}
  Here are some basic calculations of hardness for satisfiable clause-sets (for unsatisfiable $F$ see Example \ref{exp:harducls}), using Theorem \ref{thm:characuckir}:
  \begin{enumerate}
    \item $\hardness(\top) = 0$.
    \item $\hardness(\set{\set{x}}) = 0$.
    \item For $F := \set{\set{x,y},\set{x,\ol{y}}}$ we have $\hardness(F) = 1$:
      \begin{enumerate}
      \item $\primec_0(F) = \set{\set{x}}$.
      \item $\hardness(\pao{x}{0} * F) = \hardness(\set{\set{y},\set{\ol{y}}}) = 1$.
      \end{enumerate}
    \item For $F := \set{\set{\ol{x},y},\set{\ol{y},z}}$ we have $\hardness(F) = 1$:
      \begin{enumerate}
      \item $\primec_0(F) = \set{\set{\set{\ol{x},y},\set{\ol{y},z},\set{\ol{x},z}}}$.
      \item $\hardness(\pab{x \ra 1, y \ra 0} * F) = \hardness(\set{\bot}) = 0$.
      \item $\hardness(\pab{y \ra 1, z \ra 0} * F) = \hardness(\set{\bot}) = 0$.
      \item $\hardness(\pab{x \ra 1, z \ra 0} * F) = \hardness(\set{\set{y},\set{\ol{y}}}) = 1$.
      \end{enumerate}
    \item For $F := \set{\set{z,x,y},\set{z,x,\ol{y}},\set{z,\ol{x},y},\set{z,\ol{x},\ol{y}}}$ we have $\hardness(F) = 2$:
      \begin{enumerate}
      \item $\primec_0(F) = \set{\set{z}}$.
      \item $\hardness(\pao{z}{0} * F) = \hardness(\set{\set{x,y},\set{x,\ol{y}},\set{\ol{x},y},\set{\ol{x},\ol{y}}} = 2$.
      \end{enumerate}
  \end{enumerate}
\end{examp}

\section{Fundamental properties of $\Urefc_k$}
\label{sec:fundprop}

In Subsection \ref{sec:basichd} we determine hardness for various constructions. In Subsection \ref{sec:uckcontain} we consider various classes contained in some $\Urefc_k$ together with stability properties of $\Urefc_k$. Relations to alternative hierarchies from the literature are discussed in Subsection \ref{sec:Alternativehierarchies}. We conclude our discussion of basic properties of hardness in Subsection \ref{sec:Dethd}, considering the most basic cases of precise hardness-computations. We stress that (algorithmic) computation of hardness for arbitrary instances is less important here\footnote{decision of membership in $\Urefc_k$ for $k \ge 1$ is coNP-complete, as shown in Theorem \ref{thm:dethdconpcp}, which seems natural for classes with strong expressive power}, since we aim more at constructing ``soft'' (low hardness) representations than measuring hardness of given instances. What is needed is a theory to identify general constructions.

\subsection{Some basic hardness determinations}
\label{sec:basichd}

The following basic lemma follows directly by definition:
\begin{lem}\label{lem:hdvardisj}
  If two clause-sets $F$ and $F'$ are variable-disjoint, then we have:
  \begin{enumerate}
  \item If $F, F' \in \Sat$, then $\hardness(F \cup F') = \max(\hardness(F), \hardness(F'))$.
  \item If $F \in \Sat$ and $F' \in \Usat$, then $\hardness(F \cup F') = \hardness(F')$.
  \item If $F, F' \in \Usat$, then $\hardness(F \cup F') = \min(\hardness(F), \hardness(F'))$.
  \end{enumerate}
\end{lem}

Via full clause-sets $A_n$ with $n$ variables and $2^n$ clauses we obtain (unsatisfiable, simplest) examples with $\hardness(A_n) = n$, and when removing one clause for $n \ge 1$, then we obtain satisfiable examples $A_n'$ with $\hardness(A_n') = n-1$:
\begin{lem}\label{lem:hardnessfullclauseset}
  Consider a full clause-set $F \in \Cls$ (i.e., each clause contains all variables).
  \begin{enumerate}
  \item\label{lem:hardnessfullclauseset1} $\hardness(\top) = 0$.
  \item\label{lem:hardnessfullclauseset2} If $F$ is unsatisfiable then $\hardness(F) = n(F)$.
  \item\label{lem:hardnessfullclauseset3} If $F \not= \top$, then $\hardness(F) = n(F) - \min_{C \in \primec_0(F)}\abs{C}$.
  \item\label{lem:hardnessfullclauseset4} If for $F$ no two clauses are resolvable, then $\hardness(F) = 0$.
  \end{enumerate}
\end{lem}
\begin{prf}
Part \ref{lem:hardnessfullclauseset1} follows by Definition, Part \ref{lem:hardnessfullclauseset2} is Lemma 3.18 in \citep{Ku99b}, while Part \ref{lem:hardnessfullclauseset4} follows from Part \ref{lem:hardnessfullclauseset3}. It remains to show Part \ref{lem:hardnessfullclauseset3}. If $F$ is unsatisfiable, then we get Part \ref{lem:hardnessfullclauseset2}. For satisfiable $F$ and a partial assignment $\vp$ with $\var(\vp) \sse \var(F)$ it is $\vp * F$ a full clause-set with $n(\vp * F) = n(F) - n(\vp)$, and so the assertion follows by reduction to the unsatisfiable case. \Qed
\end{prf}

The following lemma yields a way of pumping up hardness:
\begin{lem}\label{lem:hdp1}
  Consider $F \in \Cls$ and $v \in \Va \sm \var(F)$. Let $F' := \set{C \cup \set{v} : C \in F} \cup \set{C \cup \set{\ol{v}} : C \in F}$. Then we have $\hardness(F') = \hardness(F) + 1$.
\end{lem}
\begin{prf}
We have $\hardness(F') \le \hardness(F)+1$ by definition (if $v$ is not set by the test-assignment, then it can be set to an arbitrary value, yielding a forced assignment at level $\hardness(F)$). Now consider a partial assignment $\vp$ with $\var(\vp) \sse \var(F)$, $\vp * F \in \Usat$ and $\hardness(\vp * F) = \hardness(F)$. Now also $\vp * F' \in \Usat$ holds, where $\vp * F' = \set{C \cup \set{v} : C \in \vp * F} \cup \set{C \cup \set{\ol{v}} : C \in \vp * F}$. Thus we have reduced the assertion of the lemma to the special case where $F \in \Usat$, and where $\hardness(F') \ge \hardness(F) + 1$ is left to be shown. This now follows easily by induction on the number of variables. \Qed
\end{prf}

\subsection{Containment and stability properties}
\label{sec:uckcontain}

The following fundamental lemma is obvious from the definition:
\begin{lem}\label{lem:basichd}
  Consider $\mc{C} \sse \Cls$ stable under application of partial assignments and $k \in \NNZ$. If $\mc{C} \cap \Usat \sse \Urefc_k$ then $\mc{C} \sse \Urefc_k$.
\end{lem}

We apply Lemma \ref{lem:basichd} to various well-known classes $\mc{C}$ (stating in brackets the source for the bound on the unsatisfiable cases).
\begin{lem}\label{lem:hdupperb}
  Consider $F \in \Cls$.
  \begin{enumerate}
  \item\label{lem:hdupperb1} For $\vp \in \Pass$ we have $\hardness(\vp * F) \le \hardness(F)$ (by Lemma 3.11 in \citep{Ku99b}).
  \item\label{lem:hdupperb2} $\hardness(F) \le n(F)$ (by Lemma 3.18 in \citep{Ku99b}).
  \item\label{lem:hdupperb3} If $F \in \Pcls{2} = \set{F \in \Cls \mb \fa\, C \in F : \abs{C} \le 2}$, then $\hardness(F) \le 2$ (by Lemma 5.6 in \citep{Ku99b}).
  \item\label{lem:hdupperb4} If $F \in \Ho = \set{F \in \Cls \mb \fa\, C \in F : \abs{C \cap \Va} \le 1}$ (Horn clause-sets), then $\hardness(F) \le 1$ by (Lemma 5.8 in \citep{Ku99b}).
  \item\label{lem:hdupperb5} More generally, if $F \in \Qho$, the set of q-Horn clause-sets (see Section 6.10.2 in \cite{CramaHammer2011BooleanFunctions}, and \cite{Ma99j}), then $\hardness(F) \le 2$ (by Lemma 5.12 in \citep{Ku99b}).
  \item\label{lem:hdupperb6} Generalising Horn clause-sets to the hierarchy $\Ho_k$ from \cite{Kl93} (with $\Ho_1 = \Ho$): if $F \in \Ho_k$ for $k \in \NN$, then $\hardness(F) \le k$ (by Lemma 5.10 in \citep{Ku99b}).
  \end{enumerate}
\end{lem}
Obviously Part \ref{lem:hdupperb4} of Lemma \ref{lem:hdupperb} can be generalised to $F \in \Rho$ (see Lemma \ref{lem:stabuc}, Part \ref{lem:stabuc3}). And considering Part \ref{lem:hdupperb3}, by a standard autarky-argument for $\Pcls{2}$ (see \citep{Kullmann2007HandbuchMU}) we can sharpen the hardness-upper-bound $2$ for \emph{satisfiable} clause-sets:
\begin{lem}\label{lem:2CNFphard}
  For $F \in \Pcls{2} \cap \Sat$ we have $\hardness(F) \le 1$.
\end{lem}
\begin{prf}
Consider a partial assignment $\vp$ with unsatisfiable $\vp * F$. Now we have $\rk_1(\vp * F) = \set{\bot}$, since otherwise $\rk_1(\vp * F) \sse F$, and thus $\rk_1(\vp * F)$ would be satisfiable. \Qed
\end{prf}

We have the following stability properties:
\begin{lem}\label{lem:stabuc}
  Consider $k \in \NNZ$.
  \begin{enumerate}
  \item\label{lem:stabuc1} $\Urefc_k$ is stable under application of partial assignments (with Lemma \ref{lem:hdupperb}, Part \ref{lem:hdupperb1}; this might reduce hardness).
  \item\label{lem:stabuc2} $\Urefc_k$ is stable under variable-disjoint union (with Lemma \ref{lem:hdvardisj}).
  \item\label{lem:stabuc3} $\Urefc_k$ is stable under renaming variables and switching polarities (by definition).
  \item\label{lem:stabuc4} $\Urefc_k$ is stable under subsumption-elimination (by basic properties of resolution).
  \item\label{lem:stabuc5} $\Urefc_k$ is stable under addition of inferred clauses (by definition; this might reduce hardness).
  \end{enumerate}
\end{lem}

\begin{examp}\label{exp:nonstabuc}\sloppy
  Examples for non-stability:
  \begin{enumerate}
  \item $\Urefc_0$ is obviously not stable under removal of clauses.
  \item $\Urefc_0$ is not stable under removal of literal occurrences, for example $\set{\set{x,y},\set{\ol{x},\ol{y}}} \in \Urefc_0$, but $\set{\set{x},\set{\ol{x},\ol{y}}} \notin \Urefc_0$.
  \item $\Urefc_0$ is not stable under crossing out of variables, e.g.\ $\set{\set{x,y},\set{\ol{x},\ol{y}}} \in \Urefc_0$, but when crossing out variable $x$ we obtain $\set{\set{y},\set{\ol{y}}} \notin \Urefc_0$.
  \item $\Urefc_0$ is not stable under addition of clauses, for example $\set{\set{x}} \in \Urefc_0$, but $\set{\set{x},\set{\ol{x}}} \notin \Urefc_0$.
  \item $\Urefc_0$ is not stable under addition of literal occurrences, e.g.\ $\set{\set{x},\set{y}} \in \Urefc_0$, but $\set{\set{x,\ol{y}},\set{y}} \notin \Urefc_0$.
  \end{enumerate}
\end{examp}

\subsection{Alternative hierarchies}
\label{sec:Alternativehierarchies}

No class $\Urefc_k$ is stable under removal of clauses. We will see in this subsection that this boils down to the class $\mc{U}_0$ of clause-sets containing the empty clauses not being stable under removal of clauses. Some classes contained in $\Urefc_1$ however are stable under removal of clauses, for examples renamble Horn clause-sets ($\Rho$), and in \cite{CepekKucera2005GHorn} hierarchies based on this more restricted class have been considered. To understand the connection to our approach, some comments on the use of ``oracles'' in this setting are needed (see Subsection \ref{sec:outlinerelhd} for future developments).

In \citep{Ku99b,Ku00g} the hierarchy $G_k(\mc{U},\mc{S}) \sse \Cls$ ($k \in \NNZ$) has been introduced, using oracles $\mc{U} \sse \Usat$ for unsatisfiability detection and $\mc{S} \sse \Sat$ for satisfiability detection:
\begin{enumerate}
\item The minimal oracles considered there are $\mc{U}_0 := \set{F \in \Cls : \bot \in F}$ and $\mc{S}_0 := \set{\top}$.
\item One uses $G_k^0(\mc{U},\mc{S}) := G_k(\mc{U},\mc{S}) \cap \Usat$ and $G_k^1(\mc{U},\mc{S}) := G_k(\mc{U},\mc{S}) \cap \Sat$. Since $G_k^0(\mc{U},\mc{S})$ does not depend on $\mc{S}$, one writes $G_k^0(\mc{U}) := G_k^0(\mc{U},\mc{S})$.
\item For all $k \in \NNZ$ holds $G_k^0(\mc{U}_0) = \Urefc_k \cap \Usat$. On satisfiable instances in general the hierarchies are incomparable.
\item If $\mc{C} \sse \Cls$ is stable under application of partial assignments, then each class $G_k(\mc{C}) := G_k(\mc{C} \cap \Usat, \mc{C} \cap \Sat)$ (for $k \in \NNZ$) is also stable under partial assignments (Lemma 4.2 in \citep{Ku00g}). So if $\mc{C} \cap \Usat \sse \Urefc_{k'}$ for some $k' \in \NNZ$, then we have $G_k(\mc{C}) \sse \Urefc_{k+k'}$ (using Lemma \ref{lem:basichd}). This is the basis of all inclusion-relations of Section \ref{sec:fundprop}.
\item In \citep{Ku99b,Ku00g} it is assumed that $\mc{U}_0 \sse \mc{U}$ holds. This ensures that $\Urefc_k \cap \Usat \sse G_k^0(\mc{C})$ always holds, but in most cases makes classes $G_k(\mc{U},\mc{S})$ unstable under elimination of clauses.
\end{enumerate}

In \cite{CepekKucera2005GHorn} two hierarchies $(\Pi_k)_{k \in \NNZ}$, $(\Upsilon_k)_{k \in \NNZ}$ have been introduced; the basic motivations and the relations to our hierarchies are as follows:
\begin{enumerate}
\item We have $\Pi_k \cap \Usat = G_k^0(\Rho)$ and $\Pi_k \cap \Sat \sse G_k^1(\Rho)$ (with $\Pi_0 = \Rho$). Note that we do not have $\mc{U}_0 \sse \Rho$ here.
\item It is $\Rho \cap \Usat \subset G_1^0(\mc{U}_0)$ (Lemma \ref{lem:hdupperb}, Part \ref{lem:hdupperb4}), while $\Rho \cap \Sat$ is not included in any $G_k^1(\mc{U}, \mc{S}_0)$. More generally we have $\Pi_k \cap \Usat \subset G_{k+1}^0(\mc{U}_0)$ for all $k \ge 0$.
\item So the choice of the oracle $\Rho$ is less powerful on unsatisfiable instances than the choice of $\mc{U}_0$ (when going up one level in the hierarchy), while the special recognition of satisfiability for $\Rho$ is (naturally) not captured by any level of the $G_k$-hierarchy, when using only the trivial satisfiability-oracle $\mc{S}_0$ (even using $\mc{U} = \Usat$ does not change this, since this only yields full handling of all \emph{forced} assignments, while a satisfiable instance in $\Rho$ might not have any forced assignment).
\item For $k \ge 1$ we have $\Pi_k \cap \Sat \subset G_k^1(\Rho)$, where an example for $F \in G_k^1(\Rho) \sm \Pi_k$ is given by $F := \set{\set{v} \cup C : C \in F'}$ for some $F' \in \Cls \sm \Pi_k$ and $v \in \Va \sm \var(F')$. The point is that recognition for the $G_k(\mc{U},\mc{S})$-hierarchy already includes satisfiability-decision (at lower levels), and if one branch, here $\pao v1$, yields a satisfiable instance, then the other branch ($\pao v0$) is not inspected --- which however is the case for $\Pi_k$.
\item $\Rho$ is stable under application of partial assignments, and, that is its main feature, stable under removal of clauses. This yields that all $\Pi_k$ are stable under removal of clauses, which is the main motivation for this choice of the base oracle.
\item $\mc{U}_0$ is not contained in any $\Pi_k$, and thus there are unsatisfiable clause-sets of hardness $0$ not contained in any given $\Pi_k$.
\item \cite{CepekKucera2005GHorn} considered also (shortly) the hierarchy $\Upsilon_k \subset \Cls$ ($k \in \NNZ$), with $\Upsilon_k \cap \Usat = G_k^0(\Qho)$ and $\Upsilon_k \cap \Sat \sse G_k^1(\Qho)$, based on the stronger oracle $\Qho \supset \Rho$ of q-Horn clause-sets (again stable under application of partial assignments and removal of clauses). We have $\Upsilon_k \cap \Usat \subset G_{k+2}^0(\mc{U}_0)$ for all $k \ge 0$ (Lemma \ref{lem:hdupperb}, Part \ref{lem:hdupperb5}).
\end{enumerate}
By Lemma \ref{lem:basichd} we get:
\begin{lem}\label{lem:piupshd}
  For all $k \in \NNZ$ we have $\Pi_k \subset \Urefc_{k+1}$ and $\Upsilon_k \subset \Urefc_{k+2}$ for the hierarchies $\Pi_k, \Upsilon_k$ introduced in \cite{CepekKucera2005GHorn}.
\end{lem}

\subsection{Determining hardness computationally}
\label{sec:Dethd}

By the well-known computation of $\primec_0(F)$ via resolution-closure we obtain:
\begin{lem}\label{lem:dethd0}
  Whether for $F \in \Cls$ we have $\hardness(F) = 0$ or not can be decided in polynomial time, namely $\hardness(F) = 0$ holds if and only if $F$ is stable under resolution modulo subsumption (which means that for all resolvable $C, D \in F$ with resolvent $R$ there exists $E \in F$ with $E \sse R$).
\end{lem}
Thus if the hardness is known to be at most $1$, we can compute it efficiently:
\begin{corol}\label{cor:dethdhorn}
  Consider a class $\mc{C} \sse \Cls$ of clause-sets where $\mc{C} \sse \Urefc_1$ is known. Then for $F \in \mc{C}$ one can compute $\hardness(F) \in \set{0,1}$ in polynomial time.
\end{corol}
Examples for $\mc{C}$ are given by $\Ho \subset \Urefc_1$ (Lemma \ref{lem:hdupperb}) and in Subsection \ref{sec:SLURalone}. Another example class with known hardness is given by $\Pcls{2} \subset \Urefc_2$ (Lemma \ref{lem:hdupperb}), and also here we can compute the hardness efficiently:
\begin{lem}\label{lem:dethd2cls}
  For $F \in \Pcls{2}$ one can compute $\hardness(F) \in \set{0,1,2}$ in polynomial time.
\end{lem}
\begin{prf}
One method is to observe that for elements of $\Pcls{2}$ the set of prime-implicates can be determined in polynomial time, while SAT-decision can be done in linear time. More efficient is the following:
\begin{enumerate}
\item Determine first whether $F$ is satisfiable or not.
\item If $F$ is satisfiable, then $\hardness(F) \in \set{0,1}$ by Lemma \ref{lem:2CNFphard}, and whether $\hardness(F) = 0$ or not can be determined by Lemma \ref{lem:dethd0}.
\item If $F$ is unsatisfiable, then it suffices to compute $\rk_0(F)$ and $\rk_1(F)$. \Qed
\end{enumerate}
\end{prf}
See Theorem \ref{thm:dethdconpcp} for coNP-completeness of determining an upper bound on hardness.

\section{The SLUR hierarchy}
\label{sec:slurhier}

We now define the $\Slur_k$ hierarchy, generalising $\Slur$ (recall Subsection \ref{sec:SLURalone}) in a natural way, by replacing $\rk_1$ with $\rk_k$. In Subsection \ref{sec:slur=uc} we show $\Slur_k = \Urefc_k$, and as application obtain coNP-completeness of membership decision for $\Urefc_k$ for $k \ge 1$. In Section \ref{sec:comphier} we determine the relations to the previous hierarchies $\Altslurstari{k}$ and $\Canoni{k}$ as discussed in Subsection \ref{sec:sluralt}.

\begin{defi}\label{def:kslur}
  Consider $k \in \NNZ$. For clause-sets $F, F' \in \Cls$ the relation \bmm{F \rkslur{k} F'} holds if there is $x \in \lit(F)$ such that $F' = \rk_k(\pao x1 * F)$ and $F' \not= \set{\bot}$. The transitive-reflexive closure is denoted by \bmm{F \rkslurs{k} F'}. The set of all fully reduced clause-sets reachable from $F$ is denoted by
  \begin{displaymath}
    \bmm{\slur_k(F)} := \set{F' \in \Cls \mb F \rkslurs{k} F' \und \neg \ex\, F'' \in \Cls : F' \rkslur{k} F''}.
  \end{displaymath}
  Finally the class of all clause-sets which are either identified by $\rk_k$ to be unsatisfiable, or where by $k$-SLUR-reduction always a satisfying assignment is found, is denoted by $\bmm{\Slur_k} := \set{F \in \Cls : \rk_k(F) \not= \set{\bot} \Ra \slur_k(F) = \set{\top}}$.
\end{defi}
We have $\Slur_1 = \Slur$ (recall Definition \ref{def:slur2}). Note also the following simple properties for $F \in \Cls$:
\begin{enumerate}
\item\label{item:kslur1} $\top \in \slur_k(F) \Lra F \in \Sat$.
\item\label{item:kslur2} For $F' \in \slur_k(F) \sm \set{\top}$ we have $F' \in \Usat$, and if $F \in \Sat$, then $\rk_k(F') \not= \set{\bot}$.
\item\label{item:kslur3} If $F \in \Slur_k$, then $F \in \Sat$ and $F \rkslurs{k} F'$ implies $F' \in \Sat$.
\end{enumerate}
Again we could define the transition relation in a less restricted way, as $F \rkslur{k} \pao x1 * F$ iff $\rk_k(\pao x1 * F) \not= \bot$, and this would yield the same class $\Slur_k$.

\begin{examp}\label{exp:kslur}
  Some examples for $\Slur_2 \sm \Slur_1$:
  \begin{enumerate}
  \item Consider the unsatisfiable clause-set $F := \set{\set{x,y},\set{x,\ol{y}},\set{\ol{x},y},\set{\ol{x},\ol{y}}}$.
    \begin{enumerate}
    \item $F \not\in \Slur_1$ because $F$ is unsatisfiable but $\rk_1(F) \not= \set{\bot}$.
    \item $F \in \Slur_2$ because $\rk_2(F) = \set{\bot}$.
    \end{enumerate}
  \item Consider the satisfiable clause-set $F' := \set{ \set{x_1,x_2} \cup C \mb C \in F }$.
    \begin{enumerate}
    \item $F' \not\in \Slur_1 = \Slur$ because $F' \rslurs F = \pab{x_1,x_2 \ra 0} * F'$, where $\slur(F) = \set{F}$ and so $F \in \slur(F')$.
    \item $F' \in \Slur_2$ because for any $\vp$ such that $F' \rkslurs{2} \vp * F'$ and $F' \not= \top$ we have one of the following two cases:
      \begin{enumerate}
      \item $\vp * F'$ is satisfiable, and so $\vp * F' \not\in \slur_2(F)$.
      \item $\vp * F'$ is unsatisfiable and so $\pab{x_1 \ra 0, x_2 \ra 0} \sse \vp$, but this contradicts the fact that $F' \rkslurs{2} \vp * F'$. That is, after setting either $x_1$ or $x_2$ to $0$, lookahead with $\rk_2$ detects unsatisfiability of $\vp * F'$ and so one can never transition to $\vp * F'$ from $F'$.
      \end{enumerate}
      Therefore $\slur_2(F') = \set{\top}$.
    \end{enumerate}
    More generally we have $\set{ \set{x_1,\dots,x_k} \cup C \mb C \in F } \in \Slur_2 \sm \Altslurstari{k}$ (recall Example \ref{exp:altslur}).
  \end{enumerate}
\end{examp}

\begin{lem}\label{lem:auxslur}
  We have for $F \in \Cls$, $k \in \NNZ$ and a partial assignment $\vp$ with $\rk_k(\vp * F) \not= \set{\bot}$ that $F \rkslurs{k} \rk_k(\vp * F)$ holds.
\end{lem}
\begin{prf}
The assignments of $\vp$ can be performed via SLUR-$k$-transitions. \Qed
\end{prf}

\subsection{SLUR = UC}
\label{sec:slur=uc}

For $F \in \Urefc_k$ there is the following polynomial-time SAT decision: $F$ is unsatisfiable iff $\rk_k(F) = \set{\bot}$. And a satisfying assignment can be found for satisfiable $F$ via self-reduction, that is, probing variables, where unsatisfiability again is checked for by means of $\rk_k$. For $k=1$ this means exactly that the nondeterministic ``SLUR''-algorithm will not fail. And that implies that $F \in \Slur$ holds, where $\Slur$ is the class of clause-sets where that algorithm never fails. So $\Urefc_1 \sse \Slur$. Now it turns out, that actually this property characterises $\Urefc_1$, that is, $\Urefc_1 = \Slur$ holds, which makes available the results on $\Slur$.

We now show that this equality between $\Urefc$ and $\Slur$ holds in full generality for the $\Urefc_k$ and $\Slur_k$ hierarchies.
\begin{thm}\label{thm:slurhdk}
  For all $k \in \NNZ$ holds $\Slur_k = \Urefc_k$.
\end{thm}
\begin{prf}
Consider $F \in \Cls$. We have to show $F \in \Slur_k \Lra \hardness(F) \le k$. For $F \in \Usat$ this follows from the definitions, and thus we assume $F \in \Sat$.

First consider $F \in \Slur_k$. Consider a partial assignment $\vp$ such that $\vp * F \in \Usat$. We have to show $\rk_k(\vp * F) = \set{\bot}$, and so assume $\rk_k(\vp * F) \not= \set{\bot}$. It follows $F \rkslurs{k} \rk_k(\vp * F)$ by Lemma \ref{lem:auxslur}, whence $\rk_k(\vp * F) \in \Sat$ contradicting $\vp * F \in \Usat$.

Now assume $\hardness(F) \le k$, and we show $F \in \Slur_k$, i.e., $\slur_k(F) = \top$. Assume there is $F' \in \slur_k(F) \sm \set{\top}$. By Property \ref{item:kslur2} for Definition \ref{def:kslur} we get $F' \in \Usat$ and $\rk_k(F') \not= \set{\bot}$. However by Lemma \ref{lem:hdupperb}, Part \ref{lem:hdupperb1} we get $\hardness(F') \le k$, and thus $\rk_k(F') = \set{\bot}$. \Qed
\end{prf}

It seemed an essential feature of the class $\Slur$, that its most natural definition is by the SLUR-algorithm; for example in \cite{FrancoSchlipf1997Report} we find the quote ``I find it interesting that the algorithm seems simpler than the conditions under which it is a decision procedure.'' By Theorem \ref{thm:slurhdk} now we have a simple characterisation of these conditions, namely that unsatisfiability after instantiation is always detected by unit-clause propagation. Using the characterisation $\Slur = \Urefc$, we can show coNP-completeness of hardness-determination:
\begin{thm}\label{thm:dethdconpcp}
  For fixed $k \in \NN$ the decision whether $\hardness(F) \le k$ (i.e., whether $F \in \Urefc_k$, or, by Theorem \ref{thm:slurhdk}, whether $F \in \Slur_k$) is coNP-complete.
\end{thm}
\begin{prf}
The decision whether $F \notin \Slur_k$ is in NP by definition of $\Slur_k$ (or use Lemma \ref{lem:hardnessgrad}). By Theorem 3 in \cite{CepekKuceraVlcek2012SLUR} we have that $\Slur$ is coNP-complete, which by Lemma \ref{lem:hdp1} can be lifted to higher $k$. \Qed
\end{prf}

\subsection{Comparison to the previous hierarchies}
\label{sec:comphier}

The alternative hierarchies $\Altslurstari{k}$ and $\Canoni{k}$ (recall Subsection \ref{sec:sluralt}) do not generalise $\rk_1$ by $\rk_k$, but extend $\rk_1$ in various ways (maintaining linear-time computation for the (non-deterministic) transitions). In this way in \cite{CepekKuceraVlcek2012SLUR,BalyoGurskyKuceraVlcek2012SLURHier} rather complicated argumentations arise, in contrast to our elegant characterisation of the classes $\Urefc_k$ in Theorem \ref{thm:characuckir}.  As a consequence, we can give short proofs that the alternative hierarchies are subsumed by our hierarchy, while already the second level of our hierarchy is (naturally) not contained in any levels of these two hierarchies (naturally, since the time-exponent for deciding whether a (non-deterministic) transition can be done w.r.t.\ hierarchy $\Slur_k$ depends on $k$). 

First we simplify and generalise the main result of \cite{BalyoGurskyKuceraVlcek2012SLURHier}, that $\Canoni{1} \sse \Slur$. By definition we have $\Canoni{0} = \Urefc_0$.
\begin{thm}\label{thm:altcanonweak}
  For all $k \in \NNZ$ we have:
  \begin{enumerate}
  \item $\Canoni{k} \sse \Urefc_k$.
  \item $\Urefc_1 \not\sse \Canoni{k}$ (and thus $\Canoni{k} \subset \Urefc_k$ for $k \ge 1$).
  \end{enumerate}
\end{thm}
\begin{prf}
By Theorem \ref{thm:characuckir} and the fact, that the Horton-Strahler number of a tree is at most the height, we see that $\Canoni{k} \sse \Urefc_k$. That $\Urefc_1 \not\sse \Canoni{k}$ can be seen by observing that there are formulas in $\Ho \cap \Usat$ with arbitrary resolution-height complexity and so $\Ho \not\sse \Canoni{k}$. By $\Ho \subset \Urefc_1$ we get $\Urefc_1 \not\sse \Canoni{k}$. \Qed
\end{prf}

Also the other hierarchy $\Altslurstari{k}$ is strictly contained in our hierarchy:
\begin{thm}\label{thm:altslurweak}
  For all $k \in \NNZ$ we have:
  \begin{enumerate}
  \item\label{lem:altslurweak1} $\Altslurstari{k} \subset \Slur_{k+1}$.
  \item\label{lem:altslurweak2} $\Slur_2 \not\sse \Altslurstari{k}$.
  \end{enumerate}
\end{thm}
\begin{prf}
Part \ref{lem:altslurweak1} follows most easily by using Lemma \ref{lem:basichd} together with the simple fact that $\slurstar(k)(F) = \set{F}$ for $F \not= \top$ implies $\rk_{k+1}(F) = \set{\bot}$; for the strictness of the inclusion use Part \ref{lem:altslurweak2}. Part \ref{lem:altslurweak2} follows from $\Canoni{2} \not\sse \Altslurstari{k}$ (Lemma 13 in \cite{BalyoGurskyKuceraVlcek2012SLURHier}), while by Theorem \ref{thm:altcanonweak} we have $\Canoni{2} \sse \Slur_2$. \Qed
\end{prf}

Part \ref{lem:altslurweak1} of Theorem \ref{thm:altslurweak} can not be improved, since $\Altslurstari{k}$ and $\Slur_{k}$ are incomparable:
\begin{lem}\label{lem:altslurincomp}
  For $k \ge 2$ holds $\Altslurstari{k} \not\sse \Slur_k$ and $\Slur_k \not\sse \Altslurstari{k}$.
\end{lem}
\begin{prf}
  That $\Slur_k \not\sse \Altslurstari{k}$ follows by Part \ref{lem:altslurweak2} of Theorem \ref{thm:altslurweak}. That $\Altslurstari{k} \not\sse \Slur_k$ follows from the fact that for the full unsatisfiable clause-set $F_k$ on $k$ variables (i.e., containing all $2^k$ clauses of length $k$) we have $F_{k+1} \in \Altslurstari{k}$ by Lemma 10 in \cite{BalyoGurskyKuceraVlcek2012SLURHier} but $F_{k+1} \not\in \Slur_k$ by Part \ref{lem:hardnessfullclauseset2} of Lemma \ref{lem:hardnessfullclauseset}. \Qed
\end{prf}

\section{Optimisation}
\label{sec:goodrpw}

We conclude by considering the question of finding, for an input-clause-set $F$, short equivalent clause-sets $F' \in \Urefc_k$ for fixed $k$. Definition \ref{def:kbaseitself} provides the appropriate notion of ``irredundancy'' via the notion of a ``$k$-base'', where irredundancy refers to both removal of literal occurrences and removal of clauses. In Theorem \ref{thm:opt2cnfw} we show that the problem is solvable in polynomial time for inputs $F \in \Pcls{2}$, while in Theorem \ref{thm:NPcomp1baseHO} we show that the problem is NP-complete even when restricting the input to Horn clause-sets with very few prime implicates.

\begin{defi}\label{def:kbaseitself}
   A clause-set $F$ is a \textbf{$k$-base} for some $k \in \nni$ if $\hardness(F) \le k$, and after removing any literal occurrence or any clause from $F$, the result $F'$ is either not equivalent to $F$ or has $\hardness(F') > k$.
\end{defi}
Remarks:
\begin{enumerate}
\item\label{def:kbaseitself1} Every $k$-base $F$ is primal, that is, $F \sse \primec_0(F)$.
\item\label{def:kbaseitself2} A clause-set $F$ is a $0$-base iff $F = \primec_0(F)$, while $F$ is an $\infty$-base iff $F$ is primal and irredundant (removal of any clause yields a clause-set not equivalent to $F$).
\item\label{def:kbaseitself3} For a given clause-set $F$, we consider the problem of computing a shortest (w.r.t.\ the number of clauses or the number of literal occurrences) equivalent $k$-base $F'$, which we call a \textbf{$k$-base for $F$}:
  \begin{enumerate}
  \item By \cite{SchaeferUmans2002PHCompletenessCompendium} for $k = \infty$ this problem is $\Sigma_2$-complete.
  \item A special case of interest here is when $F = \primec_0(F)$, in which case $F' \sse F$ must hold. Since all prime implicates are given as input, for $k < \infty$ the decision problem whether $F$ has a $k$-base of size at most $k$ ($k$ is part of the input) is now in NP. In Theorem \ref{thm:NPcomp1baseHO} we will see that this decision problem is actually NP-complete, even under rather restricted circumstances.
  \end{enumerate}
\end{enumerate}

\begin{examp}\label{exp:kbase}
  Consider the clause-set
  \begin{displaymath}
    F := \setb{\underbrace{\set{v_1,\ol{v_3},\ol{v_4}}}_{C_1},\underbrace{\set{v_2,v_3,\ol{v_4}}}_{C_2},\underbrace{\set{v_2,\ol{v_3},v_4}}_{C_3},
               \underbrace{\set{\ol{v_2},v_3,v_4}}_{C_4},\underbrace{\set{v_1,v_3,v_4}}_{C_5},\underbrace{\set{v_1,v_2}}_{C_6}}.
  \end{displaymath}
  and clause-sets $F_1 := F \sm \set{C_5}$ and $F_2 := F \sm \set{C_6}$. We have that:
  \begin{enumerate}
  \item $F$ is a $0$-base, that is, $\primec_0(F) = F$.

    We have to show that $F$ is closed under resolution modulo subsumption. We have the following possible resolutions in $F$ with the associated subsuming clauses: $C_1 \res C_2 \supset C_6$, $C_1 \res C_3 \supset C_6$, $C_2 \res C_5 \supset C_6$, $C_3 \res C_5 \supset C_6$, $C_4 \res C_6 = C_5$.
  \item $F, F_1$ and $F_2$ are the only $k$-bases ($k \in \NNZ$) that are equivalent to $F$.

    To show that there are no other $k$-bases equivalent to $F$ we must show that all other subsets of $F$ are not equivalent to F. It suffices to show that the clauses $C_1,C_2,C_3,C_4$ are irredundant (i.e., occur in all primal clause-sets equivalent to $F$) and the clause-set $F_3 := F \sm \set{C_5,C_6}$ is not equivalent to $F$. The irredundancy of $C_1,C_2,C_3,C_4$ is seen by the fact that they are not obtained as resolvents. That $F_3$ is not equivalent to $F$ follows from the fact that $F_3$ does not contain positive clauses while $F$ does.
  \item $F_1$ is a $1$-base (and $2$-base) and is equivalent to $F$ but is not a $0$-base.

    We have $C_4 \res C_6 = C_5$ and thus $F_1 \models C_5$. To see $\hardness(F_1) = 1$, observe $\hardness(\vp_{C_5} * F_1) = \hardness(\set{\set{\ol{v_2}},\set{v_2}}) = 1$.
  \item $F_2$ is a $2$-base and is equivalent to $F$ but is not a $1$-base.

    We have $(C_1 \res C_3) \res (C_2 \res C_5) = C_6$ and thus $F_2 \models C_6$. Furthermore $\hardness(\vp_{C_6} * F_2) = \hardness(\set{\set{\ol{v_3},\ol{v_4}},\set{v_3,\ol{v_4}},\set{\ol{v_3},v_4},\set{v_3,v_4}}) = 2$.
  \item Thus $F$ is neither a $1$-base nor a $2$-base.
  \end{enumerate}
\end{examp}

\begin{thm}\label{thm:opt2cnfw}
  For clause-sets $F \in \Pcls{2}$ we can compute shortest-size (minimum number of clauses or minimum number of literal occurrences) equivalent $k$-bases $F'$ for all $k \in \nni$ in polynomial time as follows:
  \begin{enumerate}
  \item If $F$ is unsatisfiable, then the best possibility is $F' := \set{\bot}$. So assume in the sequel that $F$ is satisfiable.
  \item If $F = \top$, then $F' := \top$. So assume in the sequel that $F \not= \top$.
  \item If $F$ has a forced literal $x$, then any $k$-base for $F$ contains $\set{x}$, and we can split off $x$ by considering an optimal $k$-base for $\pao x1 * F$. So we can assume w.l.o.g.\ in the sequel that $F$ has no forced literals. (Thus $F$ as well as $\primec_0(F)$ contains only clauses of length equal $2$.)
  \item Since all $k$-bases of $F$ without new variables are subsets of $\primec_0(F)$, when considering ``shortest $k$-bases'' now there is no differences between the measures $c$ (number of clauses) and $\ell$ (number of literal occurrences), and we can just speak of ``shortest $k$-bases''.
  \item The (unique) $0$-base of $F$, the set $\primec_0(F) \in \Pcls{2}$ of all prime-implicates, can be computed in polynomial time by the methods discussed in Section 5.8 in \cite{CramaHammer2011BooleanFunctions}.
  \item Every $\infty$-base of $F$ without new variables is a $1$-base (Lemma \ref{lem:2CNFphard}), and thus w.r.t.\ $k$-bases for $k \in \nni$ only the determination of shortest $1$-bases is left, where the shortest $1$-bases are precisely the smallest subsets of $\primec_0(F)$ equivalent to $F$.
  \item Finally in Chapter 9 of \cite{Chang2004HornMin} (affirmed in \cite{HemaspaandraSchnoor2011MinGenBoolFunc}) it is shown how to compute shortest equivalent sets of prime-implicates, and thus shortest $1$-bases can be computed in polynomial time.
  \end{enumerate}
\end{thm}

\begin{thm}\label{thm:NPcomp1baseHO}
  Consider $k \in \nni$.
  \begin{enumerate}
  \item\label{lem:NPcomp1baseHO1} Assume $k \ge 1$. The decision problem ``For inputs $F \in \Ho^+ \cap \Pcls{3}$ with $\primec_0(F) = F$ and $m \in \NNZ$, decide whether there is a $k$-base $F'$ of $F$ with $c(F') \le m$.'' (note that here $F' \sse F$ must hold) is NP-complete.
  \item\label{lem:NPcomp1baseHO2} For $k=0$ the decision problem ``For input $F \in \Ho$ and $m \in \NNZ$, decide whether there is a $k$-base $F'$ of $F$ with $c(F) \le m$.'' is in P.
  \end{enumerate}
\end{thm}
\begin{prf}
For Part \ref{lem:NPcomp1baseHO2} one enumerates with polynomial delay the prime implicates of $F$ (see Section 6.5 in \cite{CramaHammer2011BooleanFunctions} for efficient methods): if this process stops with at most $m$ prime implicates found, then the answer is ``yes'', otherwise the answer is ``no''.

For Part \ref{lem:NPcomp1baseHO1} we first note that the problem is in NP, since all prime clauses are given, and $\hardness(F) \le 1$. The heart of the completeness is Theorem 6.18 in \cite{CramaHammer2011BooleanFunctions}, which states that ``Horn minimisation w.r.t.\ the number of clauses remains NP-complete even if the input is restricted to cubic pure Horn expressions.'', plus the fact from the underlying report \cite{BorosCepek1994HornNP}, that for the considered $G \in \Ho^+ \cap \Pcls{3}$ all prime implicates are also of length at most $3$, and thus we can take as input $F := \primec_0(G) \in \Ho^+ \cap \Pcls{3}$ (which can be computed in polynomial time). \Qed
\end{prf}

\section{Conclusion and outlook}
\label{sec:conclusion}

We brought together two streams of research, one started by \cite{Val1994UnitResolutionComplete} in 1994, introducing $\Urefc$ for knowledge compilation, and one started by \cite{SAFS95} in 1995, introducing $\Slur$ for polytime SAT decision. Two natural generalisations, $\Urefc_k$ and $\Slur_k$ have been provided, and the (actually surprising) identity $\Slur_k = \Urefc_k$ provides both sides of the equation with additional tools. Various basic lemmas have been shown, providing a framework for elegant and powerful proofs. Regarding computational problems, we solved the most basic questions.

Our main future application, which brings the $\Urefc$-perspective and the $\Slur$-perspective together, is in the area of ``good SAT representations''; see Subsection \ref{sec:outlinegoodrep} for more information. We consider the approach of representing a boolean function $f$ via a clause-set $F \in \Urefc_k$ as the first beginning of what we envisage as a theory of good SAT representations.

We outline now what seems to us the most promising directions for future investigations (and where we already have partial results).

\subsection{Propagation-hardness}
\label{sec:outlinephard}

Complementary to ``unit-refutation completeness'' there is the notion of ``propagation completeness'', as investigated in \cite{DarwichePipatsrisawat2011ClauseLearnRes,BordeauxMarquesSilva2012KnowledgeCompilation}. This will be captured and generalised by a corresponding measure $\phardness: \Cls \ra \NNZ$ of ``propagation-hardness'', defined as follows:
\begin{defi}\label{def:phardness}
  For $F \in \Cls$ we define the \textbf{propagation-hardness} (for short ``p-hardness'') $\bmm{\phardness(F)} \in \NNZ$ as the minimal $k \in \NNZ$ such that for all partial assignments $\vp \in \Pass$ we have
  \begin{displaymath}
    \rk_k(\vp * F) = \rki(\vp * F).
  \end{displaymath}
\end{defi}
Now the class $\Propc$ of ``propagation-complete clause-sets'' can be properly generalised:
\begin{defi}\label{def:pc}
  For $k \in \NNZ$ let $\bmm{\Propc_k} := \set{F \in \Cls : \phardness(F) \le k}$ (the class of \textbf{propagation-complete clause-sets of level $k$}).
\end{defi}
We have $\Propc = \Propc_1$. These classes lie (strictly) between the $\Urefc_k$-classes:
\begin{lem}\label{lem:hierphd}
  For $k \in \NNZ$ we have $\Propc_k \subset \Urefc_k \subset \Propc_{k+1}$.
\end{lem}

\subsection{Good representations of boolean functions}
\label{sec:outlinegoodrep}

The real power of SAT representations comes with \textbf{new variables}. Expressive power and limitations of ``good representations'' have to be studied. In the SAT-context the most useful notion of ``representation'' of a boolean function $f$ seems to be $\Sigma_1$-QCNF-representations, that is, clause-sets $F$ with $\var(f) \sse \var(F)$, where the new variables (in $\var(F) \sm \var(f)$) are implicitly existentially quantified --- in other words, the satisfying assignments of $F$ projected to the variables of $f$ are precisely the satisfying assignments of $f$; see \cite{BubeckBuening2010Definitions} for some general results. The restricted representations we already considered in Subsection \ref{sec:introoutlook} are those without new variables, that is, where $\var(F) = \var(f)$.

Additional conditions on $F$ are needed to get ``effective'' representations, since in general the evaluation of $F$ for a total assignment for $f$ is an NP-problem. Strong representations are those with bounded hardness. Strengthening Conjecture \ref{con:hierarchygoodrepw} from the introduction, we conjecture that also with new variables the power of representing boolean functions increases when allowing higher hardness:
\begin{conj}\label{con:nocollapseabshd}
  For every $k \in \NNZ$ the set of sequences $(f_n)_{n \in \NN}$ of boolean functions having sequences $(F_n)_{n \in \NN}$ of polysize-representations of p-hardness at most $k$ (i.e., $\phardness(F_n) \le k$ for all $n$) is strictly smaller then those having polysize-representations of hardness at most $k$ (i.e., $\hardness(F_n) \le k$ for all $n$), which in turn is strictly smaller then those having polysize-representations of p-hardness at most $k+1$ (i.e., $\phardness(F_n) \le k+1$ for all $n$).
\end{conj}
We wish to remind the reader of the open problem mentioned in Subsection \ref{sec:Schaeffer} about the existence of a polysize-representation of bounded hardness for affine boolean functions.

We need to emphasise here that representations $F$ of boolean functions $f$ with $\hardness(F) \le k$ fulfil an \textbf{absolute condition}, that is, we can determine unsatisfiability by $\rk_k$ for \emph{arbitrary} partial assignments, not just those using only the variables of $f$. When only asking for this \textbf{relative condition} (currently the standard, posing conditions only on variables occurring in the represented boolean function $f$, ignoring the new variables of $F$), then by generalising \cite{BKNW2009CircuitComplexity} we can show that the hierarchies collapse to the first level. This is due to the ``uncontrolled'' use of the new variables (the relative condition doesn't pose conditions on them). See \cite{BordeauxJanotaMarquesSilvaMarquis2012UC} for a study on $\Urefc$ together with the relative condition.

\subsection{Applications to cryptanalysis}
\label{sec:outlinecrypt}

As an application of the theory of ``good representations'' we consider cryptanalytic problems, especially attacking AES/DES, as preliminary discussed in \cite{GwynneKullmann2011TranslationsPrelim,GwynneKullmann2011HardnessPrelim}. For the experimental evaluation we consider the various boolean functions (``constraints'') used by these ciphers, most prominently the ``S-boxes'', and systematically search for short representations of hardness $0,1,2$ and p-hardness $1,2$. Various solvers are then run on the SAT-problems obtained by plaintext-/ciphertext pairs (where the task is to determine the key). The strengthened inference power seems especially interesting for the combination of look-ahead (``tree-resolution based'') and conflict-driven (``dag-resolution based'') SAT solvers as introduced in \cite{HeuleKullmannWieringaBiere2011Cubism}.

\subsection{Relativised hardness}
\label{sec:outlinerelhd}

Generalising \cite{BKNW2009CircuitComplexity} we can show that for example the satisfiable pigeonhole formulas $\php^m_m$ do not have polysize representations of bounded hardness even for the relative condition. One way to overcome this barrier is to generalise the theory started here via the use of oracles as in \citep{Ku99b,Ku00g} (recall Subsection \ref{sec:Alternativehierarchies}), and then employing oracles which can handle pigeonhole formulas. The basic definitions are as follows.
\begin{defi}\label{def:rku}
  A \textbf{valid oracle} for generalised unit-clause propagation is some $\mc{U} \sse \Usat$ with $\set{\bot} \in \mc{U}$ which is stable under application of partial assignments. The oracle is \textbf{strong} if $\mc{U}_0 \sse \mc{U}$, where $\bmm{\mc{U}_0} := \set{F \in \Cls : \bot \in F}$.

  Consider $k \in \NNZ$. In \citep{Ku99b} the reduction $\rk_k^{\mc{U}}: \Cls \ra \Cls$ has been defined. An equivalent definition (generalising Definition \ref{def:rk}) is as follows for $F \in \Cls$:
  \begin{eqnarray*}
    \rk_0^{\mc{U}}(F) & := &
    \begin{cases}
      \set{\bot} & \text{if } F \in \mc{U}\\
      F & \text{otherwise}
    \end{cases}\\
    \rk_{k+1}^{\mc{U}}(F) & := &
    \begin{cases}
      \rk_{k+1}^{\mc{U}}(\pao x1 * F) & \text{if } \ex\, x \in \lit(F) : \rk_k^{\mc{U}}(\pao x0 * F) = \set{\bot}\\
      F & \text{otherwise}
    \end{cases}.
  \end{eqnarray*}
\end{defi}
Note $\rk_k = \rk_k^{\mc{U}_0}$. Generalising Definitions \ref{def:hardnessucls}, \ref{def:hardnesscls}:
\begin{defi}\label{def:hardnessclsU}
  Consider a valid oracle $\mc{U}$. The \textbf{hardness} $\bmm{\hardness_{\mc{U}}(F)} \in \NNZ$ (``hardness with oracle $\mc{U}$'') of an unsatisfiable $F \in \Cls$ is the minimal $k \in \NNZ$ such that $\rk_k^{\mc{U}}(F) = \set{\bot}$. And for general $F \in \Cls$ we define $\hardness_{\mc{U}}(\top) := 0$, while for $F \not= \top$ let
  \begin{displaymath}
    \hardness_{\mc{U}}(F) := \max \set{\hardness_{\mc{U}}(\vp * F) : \vp \in \Pass \und \vp * F \in \Usat} \in \NNZ.
  \end{displaymath}
\end{defi}
We have $\hardness = \hardness_{\mc{U}_0}$, and if $\mc{U}$ is strong then for all $F$ holds $\hardness_{\mc{U}}(F) \le \hardness(F)$. An interesting oracle $\mc{U}$ (with polytime membership decision) is given by the class of unsatisfiable clause-sets defined in \cite{KlerkMaarenWarners2000Relaxations} via semidefinite programming, for which we get $\hardness_{\mc{U}}(\php^m_m) = 0$.

\subsection{Width-based hardness}
\label{sec:outlinewhd}

The basic idea is to use width-restricted resolution instead of nested input resolution, in order to increase inference power from tree-resolution to dag-resolution. A basic weakness of the standard notion of width-restricted resolution, which demands that \emph{both} parent clauses must have length at most $k$ for some fixed $k \in \NNZ$ (the ``width''), is that even Horn clause-sets require unbounded width in this sense. The correct solution, as investigated and discussed in \citep{Ku99b,Ku00g}, is to use the notion of ``$k$-resolution'' as introduced in \cite{Kl93}, where only \emph{one} parent clause needs to have length at most $k$ (thus properly generalising unit-resolution).
\begin{defi}\label{def:kres}
  Consider $k \in \NNZ$.
  \begin{itemize}
  \item Two resolvable clauses $C, D$ are \textbf{$k$-resolvable} if $\abs{C} \le k$ or $\abs{D} \le k$.
  \item We use \bmm{F \vdash^k C} if there is a resolution proof $R$ of some $C' \sse C$ from $F$ such that all resolutions in $R$ are $k$-resolutions.
  \end{itemize}
\end{defi}
This allows us now to define ``width-hardness'' (accordingly the ``hardness'' only studied in this paper can be called ``tree-hardness''):
\begin{defi}\label{def:whardness}
  For $F \in \Usat$ let $\bmm{\whardness(F)} \in \NNZ$ be the minimal $k \in \NNZ$ such that $F \vdash^k \bot$ holds. And for $F \in \Cls$ let $\whardness(F) \in \NNZ$ be the minimal $k \in \NNZ$ such that for all partial assignments $\vp$ holds $\vp * F \in \Usat \Ra \vp * F \vdash^k \bot$.
\end{defi}
We have $\whardness(F) = k \Lra \hardness(F) = k$ for $k \in \set{0,1}$, while in general $\whardness(F) \le \hardness(F)$ holds (for all $F \in \Cls$).

\begin{conj}\label{con:whdhier}
   For every $k \in \NNZ$ the set of families of boolean functions having polysize representations of width-hardness at most $k$ is strictly smaller then those having polysize-representations of width-hardness at most $k+1$. For $k \ge 1$ families showing the separation can be chosen such that they have unbounded hardness.
\end{conj}
Finally we mention that, as in Subsection \ref{sec:outlinerelhd}, we also have a relativised version $\whardness_{\mc{U}}$, based on relativised $k$-resolution as studied in \citep{Ku99b,Ku00g}.

\bibliographystyle{plainnat}

\newcommand{\noopsort}[1]{}

\end{document}